\documentclass[%
 aps,
 amsmath,amssymb,
 preprint,%
 author-year,%
]{revtex4-2}

\usepackage{graphicx}
\usepackage{dcolumn}
\usepackage{bm}
\usepackage[utf8]{inputenc}
\usepackage[T1]{fontenc}
\usepackage{mathptmx}
\usepackage{etoolbox}
\usepackage{color}
\usepackage{subcaption}

\newtheorem{proposition}{Proposition}

\newtheorem{remark}{Remark}

\newcommand{\QED}{\Box} 
\newcommand{\rw}{\rightarrow} 
    
\newcommand{\Real}{\mathbb{R}}

\newcommand{\mC}{{\mathcal C}}

\newcommand{\mE}{{\mathcal E}}

\newcommand{\mO}{{\mathcal O}}

\newcommand{\mbZ}{\mathbb{Z}}

\newcommand{\md}{\mathrm{d}}

\newcommand{\beq}{\begin{equation}}
\newcommand{\eeq}{\end{equation}}
\newcommand{\beqa}{\begin{eqnarray}}
\newcommand{\eeqa}{\end{eqnarray}}
\newcommand{\nn}{\nonumber}

\newcommand{\cblack}{\textcolor{black}}

\makeatletter
\def\@email#1#2{%
 \endgroup
 \patchcmd{\titleblock@produce}
  {\frontmatter@RRAPformat}
  {\frontmatter@RRAPformat{\produce@RRAP{*#1\href{mailto:#2}{#2}}}\frontmatter@RRAPformat}
  {}{}
}%
\makeatother

  \makeatletter
    \renewcommand\@make@capt@title[2]{%
     \@ifx@empty\float@link{\@firstofone}{\expandafter\href\expandafter{\float@link}}%
      {\textbf{#1}}\@caption@fignum@sep#2\quad}%
    \makeatother
   
\makeatletter 
\renewcommand{\fnum@figure}{\textbf{Figure~\thefigure}}
\makeatother

\begin{document}


\title[Synchronization \& parameter estimation for the generalized KS equation]{A master-slave coupling scheme for synchronization and parameter estimation in the generalized Kuramoto-Sivashinsky equation}
\author{Joaqu\'{\i}n Miguez}
\affiliation{Department of Signal Theory \& Communications, Universidad Carlos III de Madrid (Spain)}
\affiliation{Instituto de Investigaci\'on Sanitaria Gregorio Mara\~n\'on (Spain)}
 
\author{Harold Molina-Bulla}
\affiliation{%
Department of Signal Theory \& Communications, Universidad Carlos III de Madrid (Spain)
}%

\author{In\'es P. Mari\~no}%
\affiliation{ 
Department of Biology \& Geology, Physics and Inorganic Chemistry, Universidad Rey Juan Carlos (Spain)}
\affiliation{Department of Women's Cancer, University College London (UK)
}%

\email{joaquin.miguez@uc3m.es, hmolina@tsc.uc3m.es, ines.perez@urjc.es}

\date{\today}

\begin{abstract}
The problem of estimating the constant parameters of the Kuramoto-Sivashinsky (KS) equation from observed data has received attention from researchers in physics, applied mathematics and statistics. This is motivated by the various physical applications of the equation and, also, because it often serves as a test model for the study of space-time pattern formation. Remarkably, most existing inference techniques rely on statistical tools, which are computationally very costly yet do not exploit the dynamical features of the system. In this paper we introduce a simple, online parameter estimation method that relies on the synchronization properties of the KS equation. In particular, we describe a master-slave setup where the slave model is driven by observations from the master system. The slave dynamics are data-driven and designed to continuously adapt the model parameters until identical synchronization with the master system is achieved. We provide a simple analysis that supports the proposed approach and also present and discuss the results of an extensive set of computer simulations. Our numerical study shows that the proposed method is computationally fast and also robust to initialization errors, observational noise and variations in the spatial resolution of the numerical scheme used to integrate the KS equation.       
\end{abstract}

\maketitle

\section{\label{sIntro}Introduction}

%
\subsection{The Kuramoto-Sivashinsky equation} 

The Kuramoto-Sivashinsky (KS) equation is a partial differential equation (PDE) that describes the spatio-temporal evolution of certain nonlinear systems. It was independently developed by Y. Kuramoto \citep{Kuramoto76,Kuramoto78}, to model waves in Belousov–Zhabotinsky reactions, and A. Sivashinsky \citep{Michelson77,Sivashinsky80} to study instabilities in laminar flame fronts. Other significant applications include the modeling of liquid films \citep{Kadry23}, long waves in viscous fluids \citep{Topper78}, drift waves in plasmas \citep{Tatsumi84} or ion-sputtered surfaces \citep{Cuerno95}.

The generalized KS equation in 1-dimensional space can be written as \citep{Kudryashov90}
\beq
u_t + uu_x + \alpha u_{xx} + \beta u_{xxx} + \gamma u_{xxxx} = 0,
\label{eqGenKS}
\eeq
where $t \in \Real$ and $x\in\Real$ are continuous time and space, respectively, $u(t,x)$ is a scalar field representing the physical magnitude of interest, $u_t = \frac{\partial u}{\partial t}$, $u_{xx}=\frac{\partial^2 u}{\partial x^2}$, $u_{xxx}=\frac{\partial^3 u}{\partial x^3}$, $u_{xxxx}=\frac{\partial^4 u}{\partial x^4}$, and the constants $\alpha, \beta, \gamma \in \Real$ determine the dynamics of the signal $u(x,t)$. This equation has received considerable attention because it displays a rich space-time behavior, including chaotic dynamics, depending on the choice of parameter values \citep{Wittenberg99,Li04,Brummitt09,Cvitanovic10,Budanur17,Kudryashov21}

%
\subsection{Parameter estimation for the KS model}

The problem of estimating the values of the constant parameters $(\alpha, \beta, \gamma)$ from observations of the signal $u(t,x)$ is of interest because of the physical applications of the KS equation and also because this model has become a testbed to assess inference techniques that can then be applied to other spatially-extended systems. 

Most parameter estimation methods designed for the KS model rely on statistical tools. Hurst {\em et al.} \cite{Hurst22} describe a maximum likelihood (ML) inference scheme that runs an ensemble of Kalman-based estimators. Huttunen {\em et al.} \cite{Huttunen18} rely on a Fourier spectral decomposition of the KS equation with periodic boundary conditions. The Fourier series is truncated to $K$ modes and an {\em ad hoc} noise term is added to the Fourier coefficients. Inference on the signal and the parameters is carried out by way of extended Kalman filtering and an importance sampling step. Lu {\em et al.} \cite{Lu17} start from a similar approach (a truncated Fourier-series representation) and then approximate the dynamics of the Fourier modes by way of an autoregressive time-series model. Martina-Perez {\em et al.} \cite{MartinaPerez21} also advocate a statistical scheme. They rely on the approximation of the PDE by way of a pre-selected set of basis functions and then tackle the Bayesian estimation of the basis dimension and the basis coefficients from the observed data.

Other methods include the use of deep learning schemes \citep{Loew20}, a joint smoothing and parameter estimation method that aims at estimating both the solution $u(t,x)$ and the parameters $(\alpha,\beta,\gamma)$ using a quasi-Newton optimization algorithm \citep{Rudy19} or a nonlinear least-squares fitting scheme that relies on the ability to simulate batches of data using a kinetic Monte Carlo procedure \citep{Hu08b}. These schemes are offline (i.e., they require the iterative batch processing of the whole set of observations) \cblack{and \citep{Loew20} requires an offline training stage before it can be applied.}

All the methodologies described above share two limitations: 
\begin{itemize}
\item they do not exploit the specific dynamics of the KS model and, as a consequence,
\item they are computationally demanding --at least when compared with typical numerical integration schemes for the KS equation. 
\end{itemize} 

An alternative approach to the parameter estimation problem that has proved successful for chaotic dynamical models relies on the ability to synchronize two suitably coupled systems. In a typical setup, one assumes that the observed data are generated by a {\em master} system with unknown parameters. The observations are used to drive a {\em slave} model that has the same form as the master system plus a coupling term that depends on the discrepancy between the two systems. The coupling is chosen in such a way that the slave model synchronizes with the master when the two systems are identical, including their parameters. Hence, the problem of parameter estimation reduces to finding the parameter values in the slave model that make the synchronization error vanish. 

The general strategy described above, in different forms, has been advocated by several authors, e.g., in \citep{Parlitz96,Maybhate99,dAnjou01,Marinho05,Marinho06,Abarbanel10,Abarbanel17,Rudy19,Pachev22,Arellano23,Atencia23}. However, it is hard to find applications of these methodologies to the KS equation. Most authors choose the Lorenz 63/96 or R\"ossler models to test their techniques \citep{Parlitz96,Maybhate99,dAnjou01,Marinho05,Marinho06,Abarbanel10,Rudy19} (let us note that the KS equation is indeed considered as a case study in \cite{Rudy19} but this is done only to show synchronization of a KS slave model with {\em known} parameters $\alpha=\gamma=1$ and $\beta=0$, \cblack{while parameter estimation is tackled by other means}). An outstanding exception is the recent work by Pachev {\em et al.} \cite{Pachev22}, who propose a synchronization-based procedure to estimate the parameters of a KS system online (i.e., concurrently with the numerical integration of the model). \cblack{The main drawback of the method in \citep{Pachev22} is that it requires the time derivatives of the observed signals, which are hard to estimate reliably in the presence of observational noise.} In addition, their method also runs a Gram-Schmidt orthogonalization procedure and solves a system of linear equations at each time step of the numerical integration scheme --which results in a significant computational cost.

Master-slave synchronization of KS systems has been specifically shown and investigated, e.g., in \cite{Tasev00,Basnarkov14}, and synchronization-based data assimilation schemes (aimed at the tracking and forecasting of the signal $u(t,x)$) have been proposed in \cite{Jardak10,Gotoda15,Penny17}. In all these references, however, the parameters of the KS model are assumed known --there is no parameter estimation.

\subsection{Contributions}

The parameter estimation method introduced in this paper relies on a master-slave setup, similar to the general schemes of \cite{Pachev22} and \cite{Marinho05,Marinho06}. Both the master and slave systems are KS models with periodic boundary conditions and they are approximated with truncated Fourier series: the master model is represented with $M$ time-varying Fourier coefficients and we construct a slave model with $K$ coefficients, $K \le M$. We assume that the state $u(t,x)$ of the master system can be observed over a spatial grid $x_0, \ldots, x_{J-1}$ and the collected data $u(t,x_j)$, $j=0, \ldots, J-1$, are used to drive the slave system via a diffusive coupling term. We prove that this configuration yields local synchronization (according to the definition in \citep{Asheghan11}) when the parameters $(\alpha,\beta,\gamma)$ are identical in the master and slave models. We also show numerically that the scheme is robust, in the sense that synchronization is attained even when the initial conditions of the master and slave model differ significantly. 

A key difference with the method in \cite{Pachev22} lies in the parameter update rule. In the scheme introduced in this paper, the parameters of the slave model evolve continuously over time according to an ODE designed to have a stationary point where identical synchronization of the master and slave models is attained. In particular, there is no need for Gram-Schmidt orthogonalization steps \cblack{and the computation of time derivatives of the observables is explicitly avoided.} The slave parameters are integrated numerically with the same procedure used to integrate the Fourier coefficients of the master and slave models. 

We present a detailed numerical assessment of the proposed synchronization-based parameter estimation scheme. In particular, we study
\begin{itemize}
\item the robustness of the method to initial errors in the slave model and the effect of the coupling strength;
\item the robustness of the synchronization scheme and the parameter estimation method to observational noise in the data collected from the master system;
\item the adaptation rate of the parameters, i.e., how quickly they converge to the master parameter values; 
\item the effect of underestimating the number of significant Fourier coefficients $M$ in the master system, i.e., the performance of the proposed scheme when $K < M$;
\item \cblack{the relative accuracy and computational cost of the proposed scheme compared to the unscented Bucy-Kalman filter; and}
\item \cblack{the performance of the parameter estimation method for different dynamical regimes (chaotic and periodic) of the master system.}
\end{itemize}

%
\subsection{Organization of the paper}

The rest of this paper is organized as follows. A master-slave scheme that guarantees local identical synchronization is described in Section \ref{sSynchro}. The parameter estimation methodology is introduced in Section \ref{sEstimation}, which also includes an extensive set of computer simulation results. Finally, Section \ref{sConclusions} is devoted to a summary and a brief discussion of the methodology.

%
\section{Synchronization of generalized Kuramoto-Sivashinsky models} \label{sSynchro}

%
\subsection{Master model} \label{ssKS}

The methods described in this paper rely on a master-slave setup. The master system is described by Eq. \eqref{eqGenKS} and we are interested on the evolution of the scalar field $u(t,x)$ during a time interval of length $T$, namely for $t \in [0,T]$. We assume a periodic boundary condition of the form $u(t,x) = u(t,x+X)$, for all $t \in [0,T]$ and some $X<\infty$, and denote the initial condition as $u_0(x) := u(0,x)$. The parameters of the master model $(\alpha,\beta,\gamma)$ are constant and assumed unknown.

The periodic-in-space field $u(t,x)$ can be equivalently represented by its time-varying Fourier series coefficients $\{ a_k(t): k\in\mbZ\}$ \citep{Christov90}. To be specific, if we let $\phi_k(x) := \exp\left\{i \frac{2\pi}{X}kx \right\}$ for $k \in \mbZ$ then we can write
\beq
u(x,t) = \sum_{k=-\infty}^\infty a_k(t) \phi_k(x),
\label{eqFSu}
\eeq 
for $t\in[0,T]$ and $x \in \Real$. If we substitute \eqref{eqFSu} into \eqref{eqGenKS}, multiply by $\phi_k^*(x)$ and integrate over $x\in[0,X)$ on both sides of the equation, then we arrive at the infinite system of nonlinear ordinary differential equations (ODEs)
\beq
\dot a_k = a_k\left(
	\alpha \omega_0^2 k^2 + i\beta \omega_0^3 k^3 - \gamma \omega_0^4 k^4
\right) - \frac{1}{2} i k \omega_0 \sum_{\ell=-\infty}^\infty a_\ell a_{k-\ell},
\label{eqFourier}
\eeq
for $k \in \mbZ$, where $\omega_0 = \frac{2\pi}{X}$. We restrict our attention to real fields, $u(t,x)\in\Real$ for all $x$ and $t\in[0,T]$, hence the Fourier coefficients are Hermitian, $a_k(t) = a_{-k}^*(t)$, and the one-sided sequence $\{a_k(t): k=0,1,2,\ldots\}$ determines the signal $u(t,x)$.

%
\subsection{Slave model} \label{ssScheme}

Assume that the master model can be observed at a grid of spatial locations $x_0, \ldots, x_{J-1} \in [0,X)$ during the time interval $t \in [0,T]$, i.e., we collect measurements $u(t,x_j)$, $j=0, \ldots, J-1$. It is possible to use these observations to estimate the Fourier coefficients of the signal $u(t,x)$. Specifically, let us assume a truncation of Eq. \eqref{eqFSu} to the first $K$ harmonics. Provided that $J \ge 2K+1$ and assuming that $u(t,x) \in \Real$ (hence $a_k(t)=a_{-k}^*(t)$), it is possible to estimate the coefficients $a_0(t), \ldots, a_{K}(t)$ by solving the linear least-squares (LS) problem
\beq
\min_{ \tilde a_k(t), k\ge 0}
\sum_{j=0}^{J-1} \left|
	u(t,x_j) - \tilde a_0(t) - \left[ 
		\sum_{k=1}^{K} \tilde a_k(t)\phi_k(x_j) + \tilde a_k^*(t)\phi_k^*(t)
	\right]
\right|^2. 
\label{eqLS}
\eeq
The solution of problem \eqref{eqLS} can be explicitly (and compactly) written as
\beq
\hat{\bm{a}}(t) := \left( \bm{\Phi}^H\bm{\Phi} \right)^{-1} \bm{\Phi}^H \bm{u}(t),
\label{eqLScoeff}
\eeq
where $(\cdot)^H$ denotes the conjugate-transpose of a matrix or vector, $\bm{u}(t) = \left[ u^*(t,x_0), \ldots, u^*(t,x_{J-1}) \right]^H$ is the $J \times 1$ vector of observations at time $t$, $\bm{\Phi} = \left[ \bm{\phi}_K^* \cdots \bm{\phi}_1^*~~\bm{\phi}_0~~\bm{\phi}_1 \cdots \bm{\phi}_{K} \right]$ is the $J \times (2K+1)$ matrix with columns given by
\beq
\bm{\phi}_k := \left[
	\begin{array}{c}
	\phi_k(x_0)\\
	\vdots\\
	\phi_k(x_{J-1})\\
	\end{array}
\right], \quad \text{and} \quad
\hat{\bm{a}}(t) = \left[
	\begin{array}{c}
	\hat a_K^*(t)\\
	\vdots\\
	\hat a_1^*(t)\\
	\hat a_0(t)\\
	\hat a_1(t)\\
	\vdots\\
	\hat a_{K}(t)\\
	\end{array}
\right]
\label{eqDef_a}
\eeq
is a $2K+1$ vector that contains the LS estimates of the Fourier coefficients of $u(t,x)$. 
\begin{remark}
The coefficients in $\hat{\bm{a}}(t)$ are {\em estimates} because we assume a truncation to order $K$ when we specify problem \eqref{eqLS}. The truncation implies that $\hat a_k(t)=0$ for $|k|>K$, while the true coefficients $a_k(t)$ can be non-zero for any $k$. 
\end{remark}
	
Given the empirical estimates in Eq. \eqref{eqLScoeff} we construct a (truncated) slave system with Fourier coefficients $b_k(t)$ that evolve over time according to the nonlinear set of ODEs
\beq
\dot b_k = b_k\left(
	\alpha \omega_0^2 k^2 + i\beta \omega_0^3 k^3 - \gamma \omega_0^4 k^4
\right) - \frac{1}{2} i k \omega_0 \sum_{\ell=-K}^K b_\ell b_{k-\ell}
+ \sum_{j=0}^{K} D_{kj} (\hat a_j - b_j), \quad k=0, \ldots, K,
\label{eqFourierSlave}
\eeq
subject to $b_k(t)=0$ for $|k|>K$ and $b_k(t) = b_{-k}^*(t)$ for every $k$. The first two terms in \eqref{eqFourierSlave} mimic the master model of Eq. \eqref{eqFourier}, and $\left\{ D_{kj} \in \mathbb{C}:~~k,j\in\{0, \ldots, K\} \right\}$ are the coefficients of a $(K+1) \times (K+1)$ coupling matrix that we denote as $\bm{D}$. We assume identical parameters $(\alpha,\beta,\gamma)$ in the master and slave models for the moment.

The field of the slave system is constructed from the Fourier modes obtained via Eq. \eqref{eqFourierSlave}, namely,
\beq
v_K(x,t) := \sum_{k=-K}^{K} b_k(t) \phi_k(x)
\nn
\eeq
for $t \in [0,T]$ and periodic boundary condition $v_K(t,x) = v_K(t,x+X)$. Since $b_k(t) = b_{-k}^*(t)$ for $k=-K, \ldots, -K$, it follows that the signal is real, i.e., $v_K(t,x) \in \Real$.

\begin{remark}
Truncation of the Fourier modes amounts to a spatial discretization. If we let $v(x,t) :=\lim_{K\rw\infty} v_K(t,x)$ be the limit field generated by the slave model, then the signal $v(t,x)$ may display small-scale spatial effects which are not present in the truncation approximation $v_K(t,x)$ if $K$ is too small. If those small-scale effects are of interest, one needs to increase $K$ in order to obtain a finer spatial resolution. 
\end{remark}

%
\subsection{Identical synchronization} \label{ssIdentical}

Assume that the master system is integrated numerically using a truncation scheme similar to the slave model. To be specific, assume that $u(t,x) \approx u_M(t,x)$, where
\beq
u_M(t,x) := \sum_{k=-M}^M \bar a_k(t) \phi_k(x),
\label{eqFirstTrunc}
\eeq
and the coefficients $\bar a_k(t)$ evolve over time according to the set of ODEs
\beq
\dot {\bar a}_k = \bar a_k\left(
	\alpha \omega_0^2 k^2 + i\beta \omega_0^3 k^3 - \gamma \omega_0^4 k^4
\right) - \frac{1}{2} i k \omega_0 \sum_{\ell=-M}^M \bar a_\ell \bar a_{k-\ell},
\label{eqFourierM},
\eeq
for $k=0, \ldots, M$, subject to $\bar a_k(t)=0$ for $|k|>M$ and $\bar a_k(t)=\bar a_{-k}^*(t)$ for all $k$, so as to guarantee that $u_M(t,x)\in\Real$. We remark that the Fourier coefficients in \eqref{eqFSu} and \eqref{eqFirstTrunc} are different, in general, for finite $K$, hence the notation $\bar a_k(t)$ for the approximation versus $a_k(t)$ for the true master system. 

We are interested in the synchronization error between the master and slave models, which we denote as
\beq
\mE(t,x) := u_M(t,x)-v_K(t,x), \quad t \in [0, T], \quad x\in \Real.
\label{eqDef_E(x,t)}
\eeq
It is relatively simple to show that, when the truncation order is the same in the two models ($K=M$) and we choose a suitable coupling matrix $\bm{D}$, $\mE(t,x) \rw 0$ for all $x$ provided that the initial conditions $u_M(0,x)$ and $v_K(0,x)$ are sufficiently close. This is the notion of {\em local synchronization} as defined, e.g., in \cite{Asheghan11}. The proposition below provides a precise statement and a formal proof of this result.   
\begin{proposition} \label{propSynchro}
Assume that the observations of the master system are $u(t,x_j)=u_M(t,x_j)$, for $j=0, \ldots, J-1$, and $J \ge 2M+1$. If 
\begin{itemize}
\item[(i)] \label{ass_i} $|u|_\infty := \sup_{x\in\Real; t \in [0,T]} |u_M(t,x)| < \infty$,
\item[(ii)] \label{ass_ii} $K=M$ and
\item[(iii)] \label{ass_iii} $\bm{D} = D \bm{I}$, where $\bm{I}$ is the identity matrix and $D\in\Real^+$,
\end{itemize}
then the synchronization error field has a stationary point at $\mE(t,x)=0$ whenever $D$ is sufficiently large.
\end{proposition}

\noindent\textit{Proof:} Since $K=M$, the observations have the form $u_K(t,x_j)$, $j=0, \ldots, J-1$. The assumption $J \ge 2M+1$ ensures that Eq. \eqref{eqLScoeff} yields the coefficients $\bar a_k$ exactly, i.e., $\hat a_k = \bar a_k$ for $k=0, \ldots, K$, with $\bar a_k=0$ when $|k|>K$ and $\bar a_k = \bar a_{-k}^*$ for all $k$.  

Let us denote $\bm{b}(t) := \left[ b_0^*(t), \ldots, b_K^*(t) \right]^H$ and $\bar{\bm{a}}(t) := \left[ \bar a_0^*(t), \ldots, \bar a_K^*(t) \right]^H$. Then, the systems of ODEs \eqref{eqFourierM} and \eqref{eqFourierSlave} yield, respectively,
\beqa
\dot {\bar{\bm{a}}} &=& \bm{\Psi}(\bm{\theta},\omega_0)\bar{\bm{a}} - \frac{1}{2} i \omega_0 \bm{\eta}(\bar{\bm{a}}), \label{eqVeca}\\
\dot{\bm{b}} &=& \bm{\Psi}(\bm{\theta},\omega_0)\bm{b} - \frac{1}{2} i \omega_0 \bm{\eta}(\bm{b}) + \bm{D}\left( \bar{\bm{a}} - \bm{b} \right),\label{eqVecb} 
\eeqa
were $\bm{\theta} = \left[ \alpha, \beta, \gamma \right]^\top$ is the $3 \times 1$ parameter vector, $\bm{\Psi}(\bm{\theta},\omega_0)$ is a $(K+1) \times (K+1)$ diagonal matrix with $k$-th diagonal entry $[\bm{\Psi}]_{kk} = \alpha \omega_0^2 k^2 + i\beta \omega_0^3 k^3 - \gamma \omega_0^4 k^4$ and $\bm{\eta}(\bm{c})$ is a $(K+1)\times 1$ vector for which the $k$-th entry is $[\bm{\eta}(\bm{c})]_k = k\sum_{\ell=-K}^K c_\ell c_{k-\ell}$.

If we define 
\beq
e_k(t) := \bar{a}_k(t)-b_k(t) 
\quad \text{and} \quad 
\bm{e}(t) := \left[ e_0^*(t), \ldots, e_K^*(t) \right]^H,
\nn
\eeq 
and then subtract \eqref{eqVecb} from \eqref{eqVeca}, we obtain the system of ODEs
\beq
\dot{\bm{e}} = \bm{\chi}(\bm{e}),
\label{eqODE_e}
\eeq
where 
$
\bm{\chi}(\bm{e}) := \left(
	\bm{\Psi}(\bm{\theta},\omega_0) - \bm{D}
\right) \bm{e} - \frac{1}{2} i \omega_0 \left(
	\bm{\eta}(\bar{\bm{a}}) - \bm{\eta}(\bar{\bm{a}}-\bm{e})
\right)
$. It is apparent from Eq. \eqref{eqDef_E(x,t)} that the synchronization error admits a Fourier series representation with coefficients $e_k(t)$, 
\beq
\mE(t,x) = \sum_{k=-K}^K e_k(t) \phi_k(t),
\label{eqFourierError}
\eeq 
hence the field $\mE(t,x)$ has a stationary point at $\mE(t,x)=0$ if, and only if, the ODE \eqref{eqODE_e} has a stationary point at $\bm{e}(t)=\bm{0}$. In order to prove that $\bm{e}(t)=0$ is a stationary point it is sufficient to show that the Jacobian of $\bm{\chi}(\bm{e})$, denoted $\bm{J}_{\chi}(\bm{e})$, is negative definite at $\bm{e}=0$, i.e., $\bm{J}_\chi(\bm{0}) \prec \bm{0}$ \citep{Regalia95}.

Some straightforward calculations show that
\beq
\bm{J}_\chi(\bm{e}) = \bm{\Psi}(\bm{\theta},\omega_0) - i\omega_0 \bm{Q}(\bar{\bm{a}},\bm{e}) - \bm{D}, \nn
\eeq
where the matrix $\bm{Q}(\bar{\bm{a}},\bm{e})$ is
\beq
\bm{Q}(\bar{\bm{a}},\bm{e}) = 
\left[
	\begin{array}{cccc}
	0 &0 &\cdots&0\\
	\bar a_1-e_1 &\bar a_0-e_0 &\cdots &\bar a_{1-K}-e_{1-K}\\
	\bar a_2-e_2 &\bar a_1-e_1 &\cdots &\bar a_{2-K}-e_{2-K}\\
	\vdots &\vdots &\ddots &\vdots\\
	\bar a_K-e_K &\bar a_{K-1}-e_{K-1} &\cdots &\bar a_0-e_0\\
	\end{array}
\right].
\nn
\eeq
Assumption (i) in the statement of Proposition \ref{propSynchro} implies that $\sup_{k,t} |\bar a_k(t)|<\infty$ which, in turn, implies that there exists some real constant $D_o<\infty$ such that, for every $D>D_o$, 
\beq
\bm{J}_\chi(\bm{0}) = \bm{\Psi}(\bm{\theta},\omega_0) - i\omega_0 \bm{Q}(\bar{\bm{a}},\bm{0}) - D\bm{I} \prec 0.
\nn
\eeq
$\QED$

\begin{remark}
The coupling matrix $\bm{D}$ can be chosen to be non-diagonal and still obtain local synchronization. Simply note that Eq. \eqref{eqODE_e} has a stationary point at $\bm{e}(t)=\bm{0}$ whenever $\bm{\Psi}(\bm{\theta},\omega_0) - i\omega_0 \bm{Q}(\bar{\bm{a}},\bm{0}) - \bm{D}$ is negative definite.
\end{remark}

\begin{remark}
We have introduced the coupling term $\bm{D}(\hat{\bm{a}}-\bm{b})$, that leads to synchronization, in the ODE of the Fourier coefficients \eqref{eqVecb}. The synchronization schemes in \cite{Tasev00,Basnarkov14,Pachev22} are similar but the coupling is introduced as an additional term directly in Eq. \eqref{eqGenKS}. 

The study of synchronization schemes for chaotic models based on a diffusive term (often called `nudging' in the context of data assimilation \cite{Abarbanel10,Pachev22}) can be traced back to the 90s, with the seminal paper by Pecora and Carroll \cite{Pecora90}.
\end{remark}

%
\subsection{Numerical results} \label{ssNumericalSynch}

Proposition \ref{propSynchro} guarantees that the slave system attains identical synchronization with the master system only when their initial conditions are sufficiently close. Our computer experiments, however, show that the scheme is robust and synchronization is achieved even when the initial conditions are significantly apart. These numerical results are discussed below.

\begin{figure}
\centerline{
	\begin{subfigure}{0.45\linewidth}
		\includegraphics[width=\linewidth]{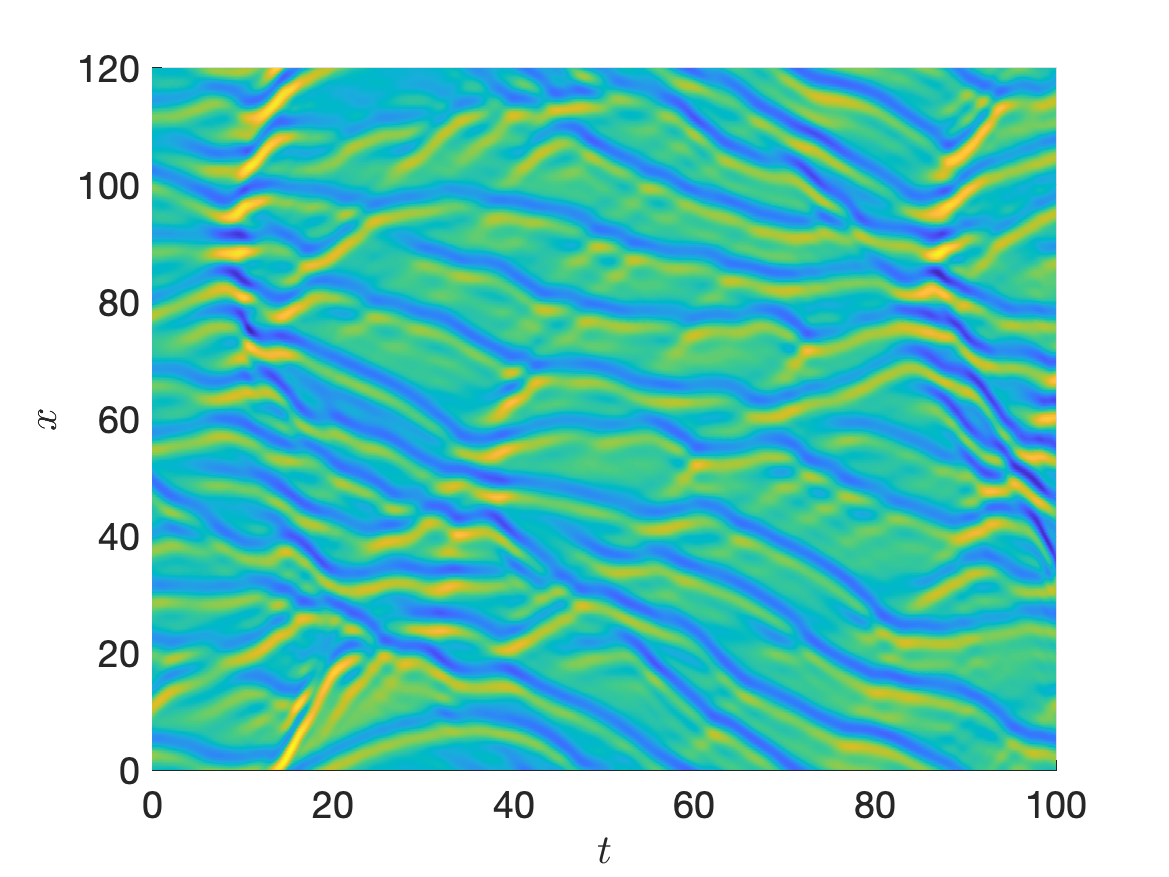}
		\caption{$u_K(t,x)$}
		\label{f1u}
	\end{subfigure}
	\begin{subfigure}{0.45\linewidth}
		\includegraphics[width=\linewidth]{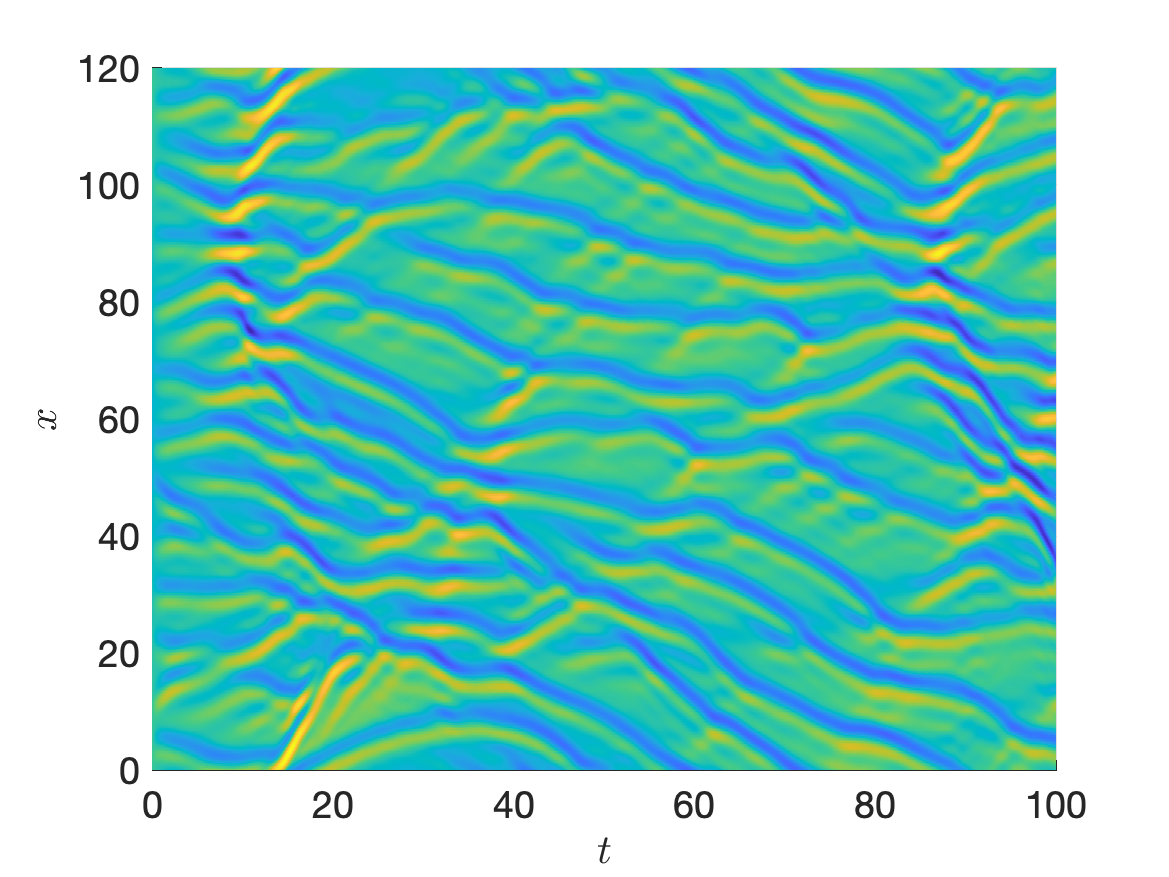}
		\caption{$v_K(t,x)$}
		\label{f1v}
	\end{subfigure}
}
\centerline{
	\begin{subfigure}{0.45\linewidth}
		\includegraphics[width=\linewidth]{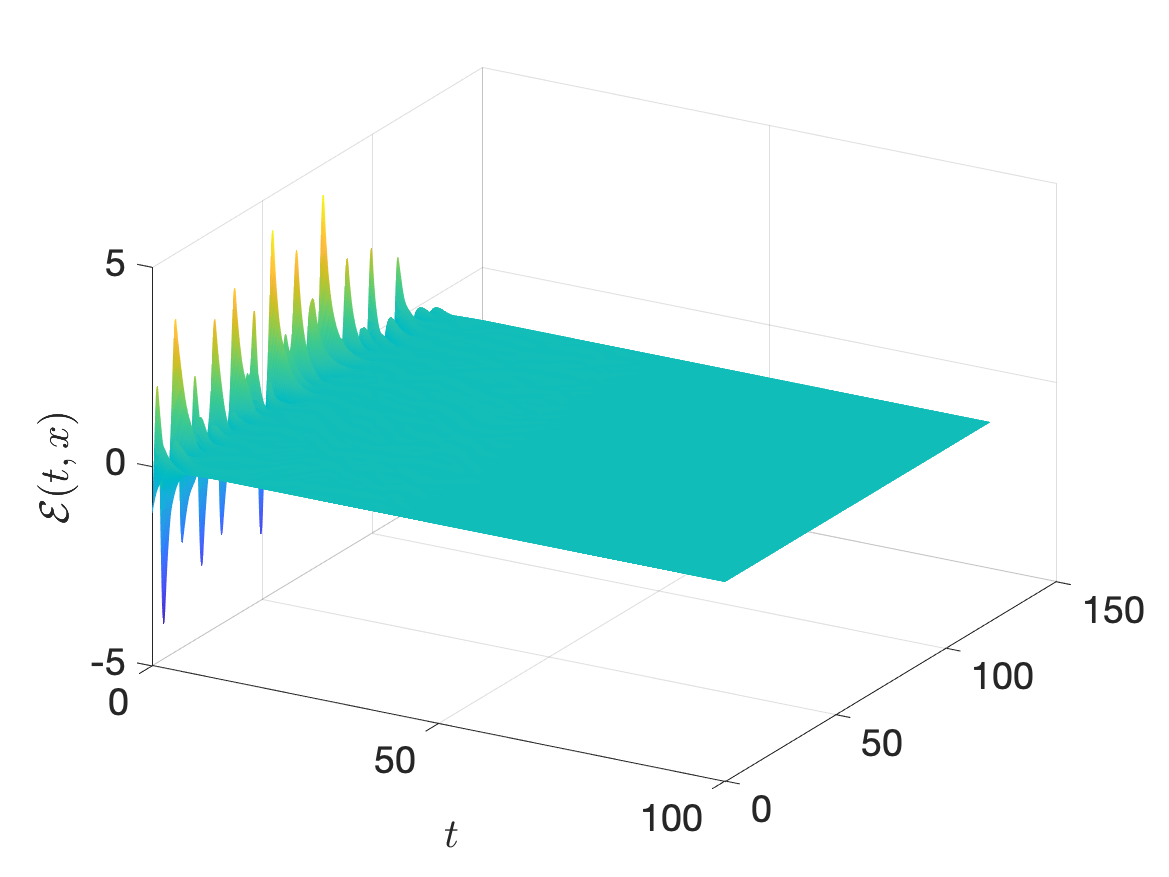}
		\caption{$\mE(t,x)=u_K(t,x)-v_K(t,x)$}
		\label{f1err}
	\end{subfigure}
}
\caption{\textbf{(a)} Real signal $u_K(t,x)$ generated by the master model, with $K=32$ Fourier modes ($2K+1=65$ coefficients), $T=100$ and $X=120$. \textbf{(b)} Real signal $v_K(t,x)$ generated by the slave model with the same values of $K$, $T$ and $X$, and a coupling constant $D=1$. \textbf{(c)} Synchronization error. We see how the large error at time $t=0$ due to the different initializations of the two systems vanishes quickly.}
\label{f1synch}
\end{figure}
	
Figure \ref{f1synch} shows the results of a computer simulation of the master and slave models with $M=K=32$ Fourier modes, a time horizon of $T=100$ time units, a spatial period $X=120$ and parameter values $\alpha=1.15$, $\beta=-0.05$ and $\gamma=0.98$. Observations are collected at positions $x_j=j$ for $j=0, \ldots, X-1$, hence $J=X=120 > 2K+1 = 65$. Numerical integration has been carried out using an Euler scheme with (sufficiently small) time step $h=0.005$, but higher order methods can obviously be used as well --the setting described in this section does not preclude the use of any numerical scheme. The coupling matrix for this simulation is $\bm{D}=D\bm{I}$ with $D=1$.  

The Fourier coefficients of the slave model are initialized as $b_{-1}(0)=b_1(0)=0.5$ and $b_k(0)=0$ otherwise. In order to have a significantly different initialization in the master model, we run a simulation of the KS equation with coefficients $\tilde a_k(t)$ and the same initialization as in the slave model, i.e., $\tilde a_{-1}(0)=\tilde a_1(0)=0.5$ and $\tilde a_k(0)=0$ for $|k|\ne 1$. Then we record the coefficient values at time $T=100$ and use them as an initial condition for the master system, i.e., $\bar a_k(0) = \tilde a_k(T)$. This procedure yields very different initial values of the master and slave Fourier coefficients and it also removes any transient regimes in the simulation of the master model.

Figure \ref{f1u} shows the signal $u_K(t,x)$ generated by the master model for $t\in[0,100]$ and $x\in[0,120)$, while Figure \ref{f1v} displays the field $v_K(t,x)$ that results from the integration of the slave model. We observe the space-time patterns typical of the KS equation and the quick synchronization of the slave system. Indeed, Figure \ref{f1err} displays the synchronization error $\mE(t,x)=u_K(t,x)-v_K(t,x)$ and we see that, even if the initial conditions of the master and slave models differ very significantly (observe the large errors at time $t=0$), $\mE(t,x)$ becomes negligible within a short time interval.

\begin{figure}
\centerline{	
	\begin{subfigure}{0.305\linewidth}
		\includegraphics[width=\linewidth]{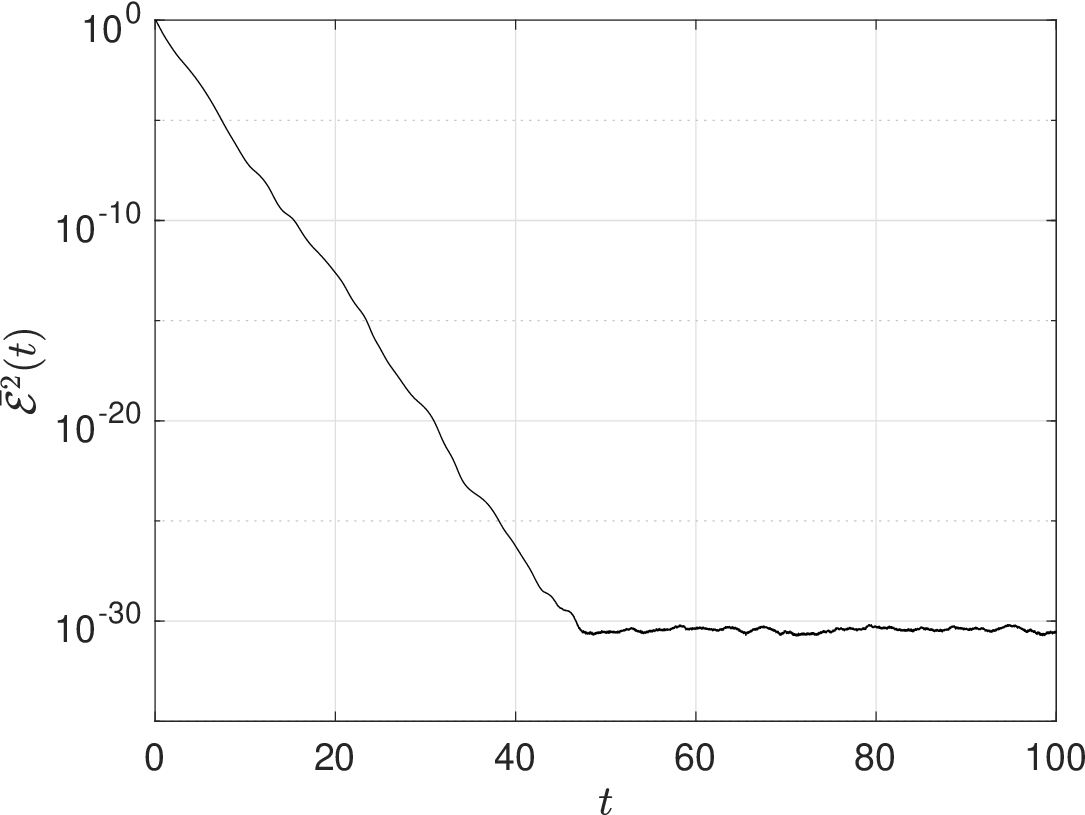}
		\caption{ }
		\label{f2mean}
	\end{subfigure}
	\begin{subfigure}{0.33\linewidth}
		\includegraphics[width=\linewidth]{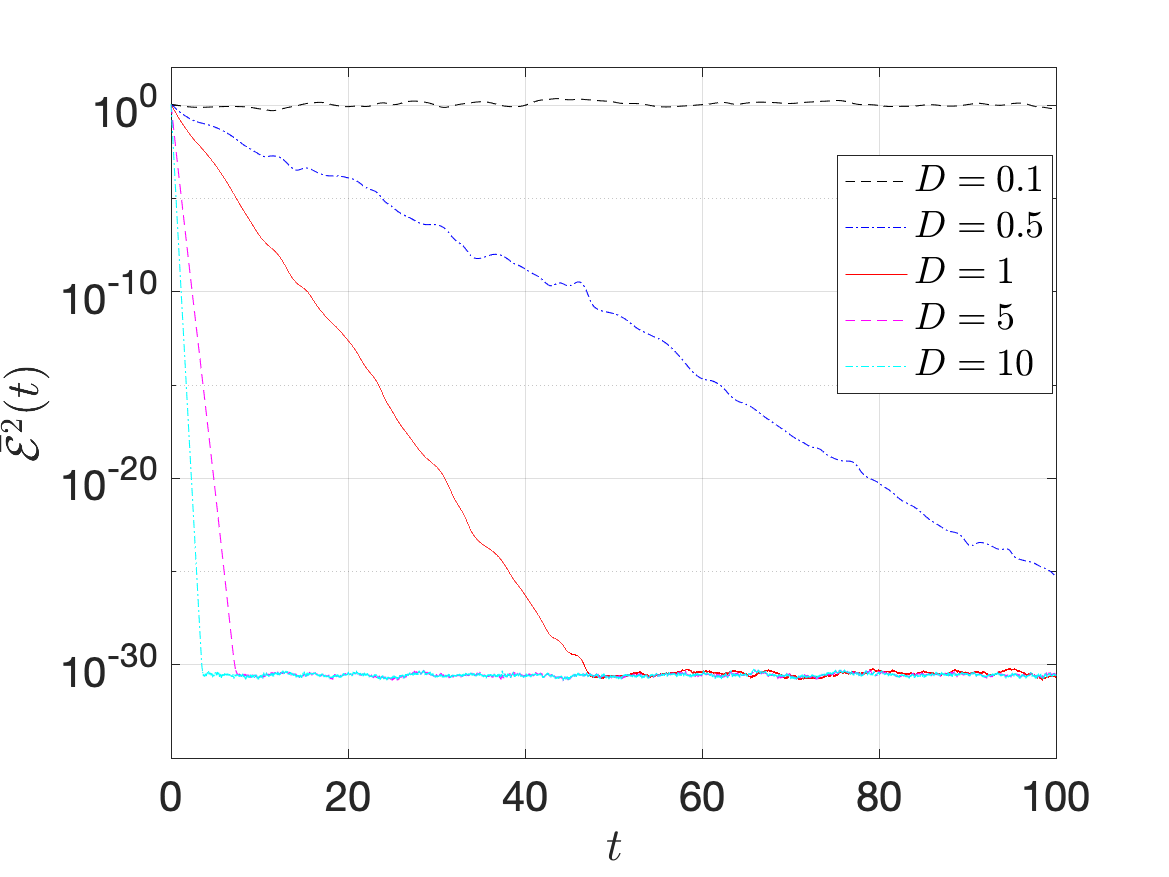}
		\caption{ }
		\label{f2norm}
	\end{subfigure}	
		\begin{subfigure}{0.33\linewidth}
		\includegraphics[width=\linewidth]{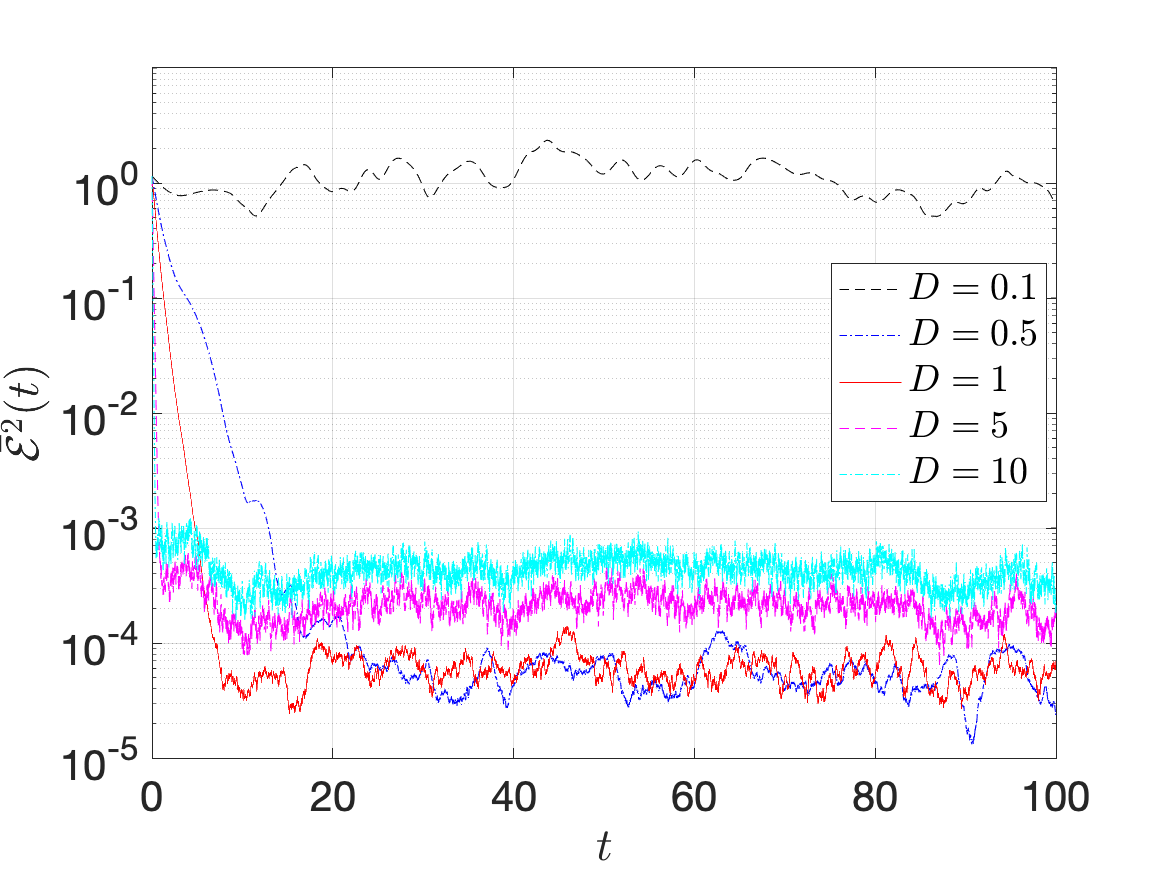}
		\caption{ }
		\label{f2noisy}
	\end{subfigure}
}
\caption{\textbf{(a)} Exponentially-fast convergence $\bar\mE^2(t)\rw 0$ of the normalized MSE. Noiseless data and $D=1$. \textbf{(b)} Convergence of the normalized MSE $\bar\mE^2(t) \rw 0$ for different choices of the coupling constant $D>0$ and noiseless data. \textbf{(c)} Convergence of the normalized MSE for different choices of $D>0$ and noisy observations. Identical parameters $(\alpha,\beta,\gamma)=(1.15, -0.05, 0.98)$ in both the master and slave models.}
\label{f2synch}
\end{figure}

Figure \ref{f2synch} enables a more accurate assessment of the synchronization error and its convergence. In particular, Figure \ref{f2mean} shows the normalized mean square error (MSE) 
\beq
\bar\mE^2(t) := \frac{\int_0^X \mE^2(t,x)\md x}{\int_0^X u_K^2(t,x)\md x},
\label{eqNormPower}
\eeq 
versus time $t$. We observe how the normalized MSE decreases exponentially fast (notice the logarithmic scale of the vertical axis) and $\bar\mE^2(t) < 10^{-10}$ already before $t=20$. 

Figure \ref{f2norm} shows again the normalized MSE $\bar\mE^2(t)$ 
versus time for different choices of the coupling coefficient $D$. In physical terms, the normalized error in \eqref{eqNormPower} is the power of the error signal relative to the power of the master signal. We observe how the coupling strength affects the synchronization rate but not the steady-state (normalized) error, which is below $10^{-30}$ for $D = 1, 5, 10$ and still converging at time $t=100$ for $D=0.5$. We also see how the coupling constant $D=0.1$ is too small to lead to synchronization.   

Finally, we assess the normalized MSE and the synchronization rate when the observations are noisy. To be specific, we run a set of simulations where the observations fed to the slave model are of the form $u_K(t,x_j) + w_j(t)$, and the noise terms $w_j(t)$ are independent white Gaussian processes with mean $0$ and power spectral density $S_w(i\omega) \approx 0.15$. 
We observe that, in the presence of noise, the slave model still synchronizes more quickly as we increase the coupling coefficient $D$. However, the steady-state normalized MSE is higher for $D \in \{5,10\}$ when compared to $D\in\{0.5, 1\}$. From this simulation, the best choice of coupling parameter is $D=1$, which yields a steady-state error $\bar\mE^2(t) \approx 10^{-4}$ already for $t > 7$.

%
\section{Parameter estimation} \label{sEstimation}

Let us assume hereafter that the parameters $\bm{\theta}=\left[\alpha,\beta,\gamma\right]^\top$ of the master model are unknown and we wish to estimate them from the observations $\bm{u}(t) = \left[ u^*(t,x_0), \ldots, u^*(t,x_{J-1}) \right]^H$ described in Section \ref{ssScheme} and the corresponding Fourier coefficients $\hat{\bm{a}}(t)$ in Eq. \eqref{eqLS}, which are collected for $t \in [0,T]$. 

The synchronization property of the KS slave model described in Section \ref{sSynchro} can be used to estimate $\bm{\theta}$. To be specific, let $\hat{\bm{\theta}} = \left[ \hat\alpha, \hat\beta, \hat\gamma \right]^\top$ be the parameters of the slave model. We have shown that, when $\hat{\bm{\theta}}=\bm{\theta}$, the slave system synchronizes and the error $\mE(t,x)$ vanishes. However, the synchronization error can be written as the Fourier series in Eq. \eqref{eqFourierError}, where the coefficients $e_k(t)=\hat a_k(t) - b_k(t)$ depend on $\hat{\bm{\theta}}$ through $b_k(t)$. Following \cite{Marinho05,Marinho06}, we argue that if the slave parameters $\hat{\bm{\theta}}$ are adapted over time in order to make $|e_k(t)| \rw 0$, then the slave model attains identical synchronization, which can only be expected when $\hat{\bm{\theta}} \rw \bm{\theta}$.   

In Section \ref{ssSlaveEstimator} we extend the slave model to account for the observation-driven adaptation of the parameters $\hat{\bm{\theta}}(t)$ over time. Then, in Section \ref{ssNumerics}, we present and discuss a set of numerical results that illustrate the validity of the method and its robustness when the observed data are noisy or the truncation order $K$ is underestimated.   

%
\subsection{Slave model} \label{ssSlaveEstimator}

We extend the slave model \eqref{eqVecb} by letting the parameters in $\bm{\theta}$ become dynamical variables. Specifically, the ODE governing the dynamics of the Fourier coefficients of the slave system is
\beq
\dot{\bm{b}} = \bm{\Psi}(\hat{\bm{\theta}},\omega_0)\bm{b} - \frac{1}{2} i \omega_0 \bm{\eta}(\bm{b}) + \bm{D}\left( \hat{\bm{a}}_\top - \bm{b} \right),
\label{eqFourierSlave2}
\eeq
where 
\beq
\hat{\bm{\theta}}(t) = \left[ 
	\begin{array}{c}
	\hat{\alpha}(t)\\
	\hat{\beta}(t)\\
	\hat{\gamma}(t)
	\end{array}
\right] 
\quad \text{and} \quad 
\hat{\bm{a}}_\top(t) = \left[ 
	\begin{array}{c}
	\hat a_0(t)\\ 
	\vdots\\
	\hat a_K(t)\\
	\end{array} 
\right]
\nn
\eeq 
is a truncated version of $\hat{\bm{a}}(t)$ in Eq. \eqref{eqDef_a} that contains the last $K+1$ entries of the original vector. Recall that $\hat a_k(t) = \bar a_k(t)$ only when we assume the master model is truncated to $M=K$ modes, otherwise (if there is no truncation or $M\ne K$) $\hat a_k(t) \ne \bar a_k(t)$ in general.

We let the time evolution of the parameters $\hat{\bm{\theta}}(t)$ be driven by the mean power of the synchronization error $\mE(t,x)$, denoted $\mE^2(t)=\frac{1}{X}\int_0^X \mE^2(t,x)\md x$. Since $\mE(t,x)$ has the Fourier series representation in Eq. \eqref{eqFourierError}, Parseval's relation yields
\beq
\mE^2(t) = \sum_{k=-K}^K |e_k(t)|^2 =\sum_{k=-K}^K |\hat a_k(t) - b_k(t)|^2 
\nn 
\eeq
which, since $\hat a_k(t) = \hat a_{-k}^*(t)$ and $b_k(t)=b_{-k}^*(t)$, can be rewritten as
\beq
\mE^2(t) = | \hat a_0(t) - b_0(t) |^2 + 2\sum_{k=1}^K |\hat a_k(t) - b_k(t) |^2.
\nn
\eeq
The complexity of the slave model can be reduced if we adopt a slightly different synchronization error function, namely,
\beq
\mC(t) := \sum_{k=0}^K |e_k(t)|^2 = \| \bm{e}(t) \|^2 = \| \hat{\bm{a}}_\top(t)- \bm{b}(t) \|^2,
\label{eqNewError}
\eeq
where $\bm{e}(t) = \hat{\bm{a}}_\top(t) - \bm{b}(t)$. The difference between $\mE^2(t)$ and $\mC(t)$ is a scale factor and the contribution of the $0$-th Fourier mode, namely,
\beq
\mC(t) = \frac{1}{2}\left(
	\mE^2(t) + |\hat a_0(t) - b_0(t)|^2
\right).
\nn
\eeq
Most importantly, $\mC(t)=\mE^2(t)=0$ if, and only if, $\hat a_k(t)=b_k(t)$ for all $k\in \{-K, \ldots, K\}$. Hence, we can adopt the cost function $\mC(t)$ to quantify the synchronization error and complete the derivation of the slave model.

Let 
$
\nabla_{\hat{\bm{\theta}}} \mC = \left[
	\frac{\partial \mC}{\partial \hat\alpha},
	\frac{\partial \mC}{\partial \hat\beta},
	\frac{\partial \mC}{\partial \hat\gamma}
\right]^\top
$
denote the gradient of the synchronization error function with respect to the parameter vector $\hat{\bm{\theta}}$. In order to progressively reduce the power of the synchronization error, we let the slave parameters evolve in the direction opposite to the gradient of $\mC$, namely,
\beq
\dot{\hat{\bm{\theta}}} = -\nabla_{\hat{\bm{\theta}}} \mC.
\label{eqGradient1}
\eeq  
Unfortunately, there is no closed-form expression for $\nabla_{\hat{\bm{\theta}}} \mC$ and, hence, it is not possible to directly use Eq. \eqref{eqGradient1}. In order to construct a practical slave system, we first approximate the error function $\mC(t)$ in terms of the parameters $\hat{\bm{\theta}}(t)$. This can be done if we linearize $\bm{b}(t)$ using a Taylor expansion around $\bm{b}(t-h)$, where $h>0$ is some small time step. From Eq. \eqref{eqVecb}, it is straightforward to see that the linearization yields
\beqa
\bm{b}(t) &=& \bm{b}(t-h) + h\dot{\bm{b}}(t-h) + \mO(h^2) 
\nn \\
&\approx& \bm{b}(t-h) + h\left(
	\bm{M}_h^H \hat{\bm{\theta}} 
	- \frac{1}{2} i \omega_0 \bm{\eta}\left(\bm{b}(t-h)\right)
	+ \bm{D}\left( 
		\hat{\bm{a}}(t-h) - \bm{b}(t-h) 
	\right)
\right),
\label{eqTaylor}
\eeqa
where
$
\bm{M}_h = \left[ \bm{m}_0 \cdots \bm{m}_K \right]
$
is a $3 \times (K+1)$ complex matrix whose $k$-th column is
\beq
\bm{m}_k = b_k^*(t-h)\left[
	\begin{array}{c}
	\omega_0^2 k^2\\
	-i \omega_0^3 k^3\\
	-\omega_0^4 k^4\\
	\end{array}
\right],
\eeq
hence, comparing with Eq. \eqref{eqVecb}, $\bm{M}_h^H \hat{\bm{\theta}} = \bm{\Psi}(\hat{\bm{\theta}},\omega_0)\bm{b}(t-h)$. 

Substituting the approximation \eqref{eqTaylor} in \eqref{eqNewError} yields, after some lengthy but straightforward calculations, an approximate gradient of the form
\beqa
\nabla_{\hat{\bm{\theta}}} \mC(t) &\approx& -h\bm{M}_h\left[
	\hat{\bm{a}}_\top(t) - \bm{b}(t-h) - h\dot{\bm{b}}(t-h)
\right]
\nn
\\
&=&-h\bm{M}_h\left[
	\hat{\bm{a}}_\top(t) - \bm{b}(t-h) 
	- h\left(
		\bm{M}_h^H\hat{\bm{\theta}} 
		- \frac{1}{2}i\omega_0 \bm{\eta}\left(
			\bm{b}(t-h)
		\right)
	\right.
\right.
\nn\\
&&\left.
	\left.
		\phantom{\frac{1}{2}}
		+ h\bm{D}\left(
			\hat{\bm{a}}_\top(t-h)-\bm{b}(t-h)
		\right)
	\right)
\right].
\label{eqGradient2}
\eeqa 
Finally, we combine approximation \eqref{eqGradient2}, for the gradient of the error, and \eqref{eqVecb}, for the Fourier coefficients $\bm{b}(t)$, to construct the slave model
\beqa
\dot{\hat{\bm{\theta}}} &=& -\mu h \bm{M}_h\left[
	\hat{\bm{a}}_\top - \bm{b}_h - h\dot{\bm{b}}_h
\right],\label{eqParUpdate}\\
\dot{\bm{b}} &=& \bm{\Psi}(\hat{\bm{\theta}},\omega_0)\bm{b} - \frac{1}{2} i \omega_0 \bm{\eta}(\bm{b}) + \bm{D}\left( \hat{\bm{a}}_\top - \bm{b} \right),
\label{eqCoeffUpdate}
\eeqa
where $\bm{b}_h=\bm{b}(t-h)$, $\dot{\bm{b}}_h=\dot{\bm{b}}(t-h)$, $\mu > 0$ is an adaptation rate parameter that can be chosen to speed up or slow down the dynamics of $\hat{\bm{\theta}}(t)$. Tuning $\mu$ and $\bm{D}$ allows for a trade-off between the time needed by the slave system to synchronize and the accuracy it achieves.

%
\subsection{Numerical results} \label{ssNumerics}

In this section we study, numerically, the ability of the slave model \eqref{eqParUpdate}-\eqref{eqCoeffUpdate} to attain synchronization with the master system \eqref{eqFSu}, and yield accurate estimates $\hat{\bm{\theta}}(t)$, when the master parameters $\bm{\theta}=[\alpha,\beta,\gamma]^\top$ are unknown.  

\begin{figure}
\centerline{
	\begin{subfigure}{0.45\linewidth}
		\includegraphics[width=\linewidth]{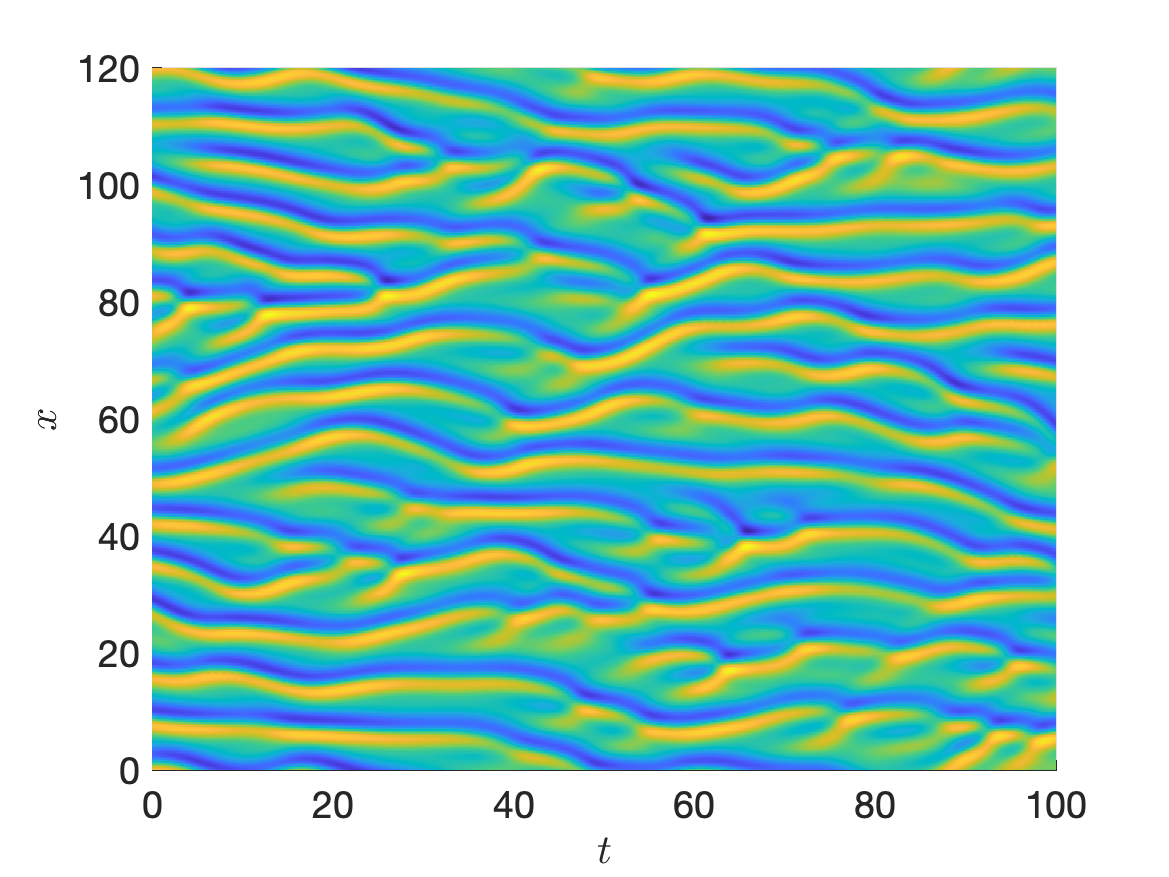}
		\caption{$u_M(t,x)$}
		\label{f3u}
	\end{subfigure}
	\begin{subfigure}{0.45\linewidth}
		\includegraphics[width=\linewidth]{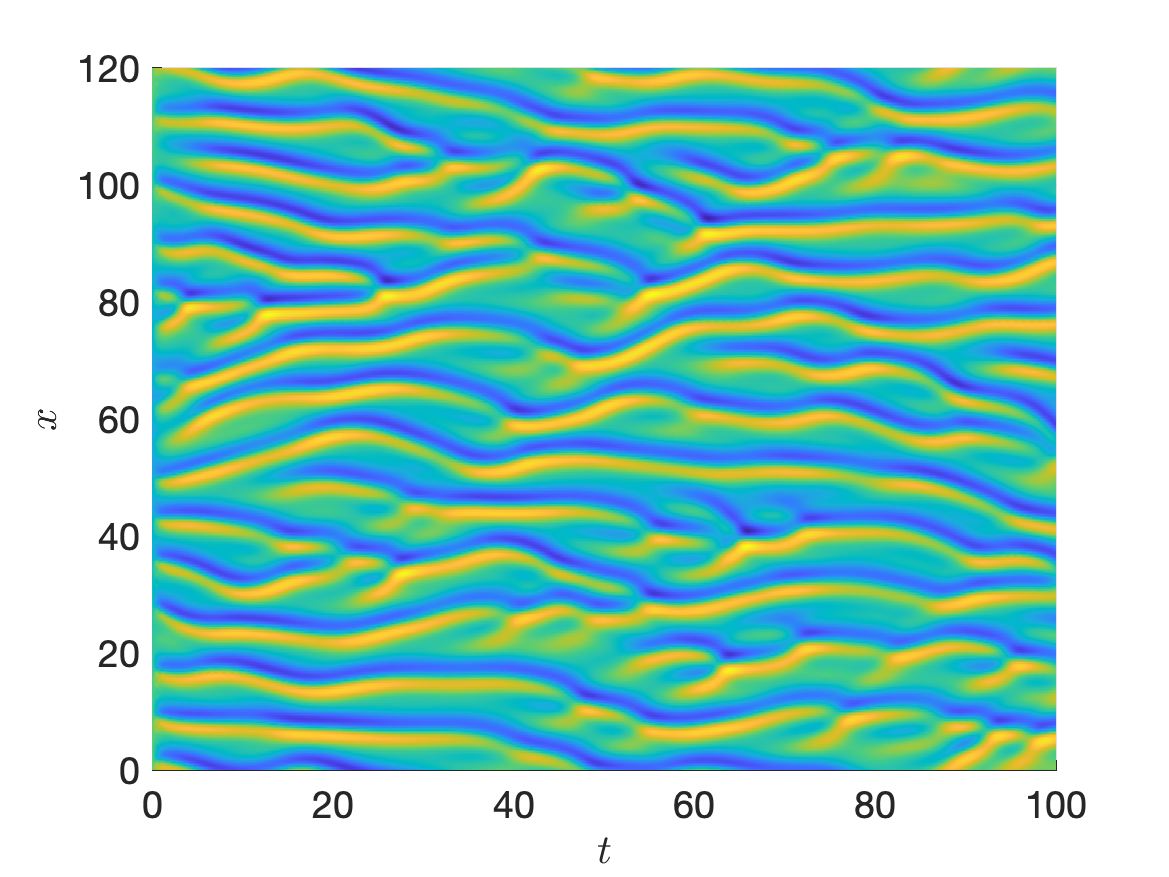}
		\caption{$v_K(t,x)$}
		\label{f3v}
	\end{subfigure}
}
\centerline{
	\begin{subfigure}{0.45\linewidth}
		\includegraphics[width=\linewidth]{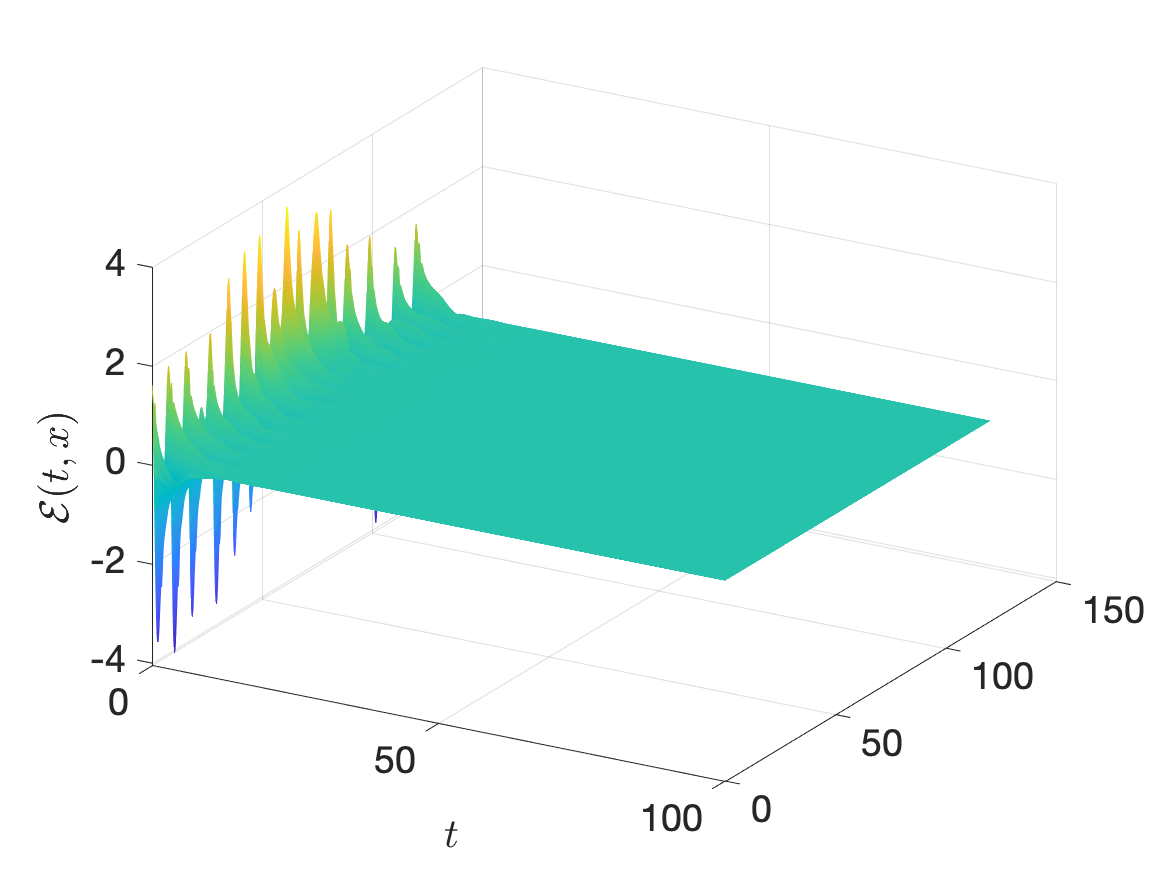}
		\caption{$\mE(t,x)=u_K(t,x)-v_K(t,x)$}
		\label{f3err}
	\end{subfigure}
}
\caption{\textbf{(a)} Real signal $u_K(t,x)$ generated by the master model, with $M=K=64$ ($2K+1=129$ Fourier coefficients), $T=100$ and $X=120$. \textbf{(b)} Real signal $v_K(t,x)$ generated by the slave model with the same values of $K$, $T$ and $X$, coupling constant $D=1$, initial parameter values $\hat{\bm{\theta}}(0)=[0,0,0]^\top$ and parameter adaptation rate $\mu=200$. \textbf{(c)} Synchronization error. We see how the large error at time $t=0$ due to the different initializations of the two systems vanishes quickly.}
\label{f3par}
\end{figure}

\subsubsection{\cblack{Simulation setup}}

The simulation setup is similar to the computer experiments in Section \ref{ssNumericalSynch}. Figure \ref{f3par} shows the results of a simulation run with $M=64$ Fourier modes in the master system and $K=64$ modes in the slave model (note that $K$ modes implies $2K+1$ coefficients, $b_{-K}(t),\ldots,b_K(t)$, however $b_k(t)=b^*_{-k}(t)$). The length of the time interval is $T=100$, the spatial period is $X=120$ and the parameter values in the master system are $\alpha=1.15$, $\beta=-0.05$ and $\gamma=0.98$, which yield chaotic dynamics. Observations are collected at positions $x_j=\frac{j}{2}$ for $j=0, 1, \ldots, 2X-1$, hence $J=2X=240 > 2K+1 = 129$. The ODEs \eqref{eqFSu} and \eqref{eqParUpdate}-\eqref{eqCoeffUpdate} are numerically integrated using an Euler scheme with (a sufficiently small) time step $h=0.005$ (but higher-order schemes can be used as well). The Fourier coefficients of the master and slave models are initialized in the same way as in Section \ref{ssNumericalSynch}, which yields a large error at time $t=0$ (see Figure \ref{f3err}). The initial parameter values at the slave system are $\hat{\bm{\theta}}(0)=[0,0,0]^\top$. The coupling matrix in \eqref{eqCoeffUpdate} is $\bm{D}=D\bm{I}$ with $D=1$ and the adaptation rate is $\mu=200$.

\subsubsection{\cblack{Synchronization and parameter estimation}}

Figure \ref{f3u} displays the field $u_K(t,x)$, $t \in [0,T]$, $x\in[0,X]$, generated by the master model, while Figure \ref{f3v} displays the signal $v_K(t,x)$ generated by the slave model. Figure \ref{f3err} shows that the synchronization error $\mE(t,x)=u_K(t,x)-v_K(t,x)$ vanishes quickly despite the significantly different initializations of the two models.

\begin{figure}
\centerline{	
	\begin{subfigure}[t]{0.45\linewidth}
		\includegraphics[width=\linewidth]{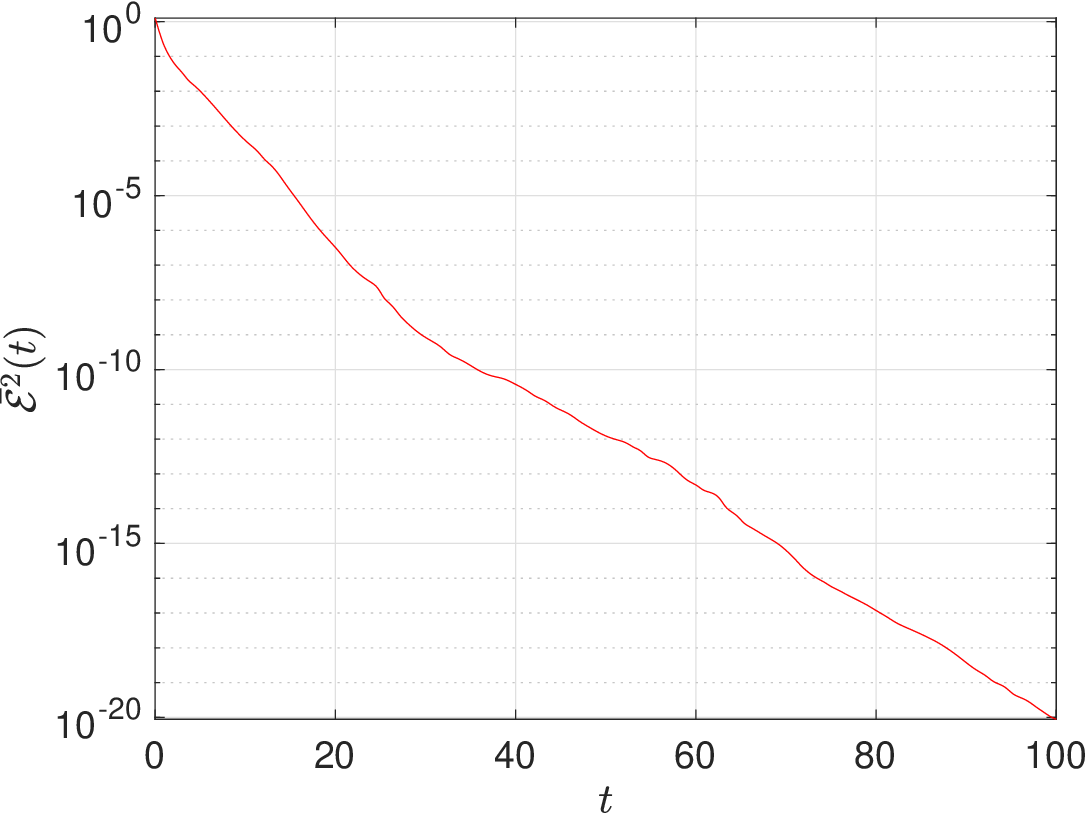}
		\caption{Normalized error $\bar{\mathcal{E}}^2(t)$.}
		\label{f4e2}
	\end{subfigure}
	\begin{subfigure}[t]{0.432\linewidth}
		\includegraphics[width=\linewidth]{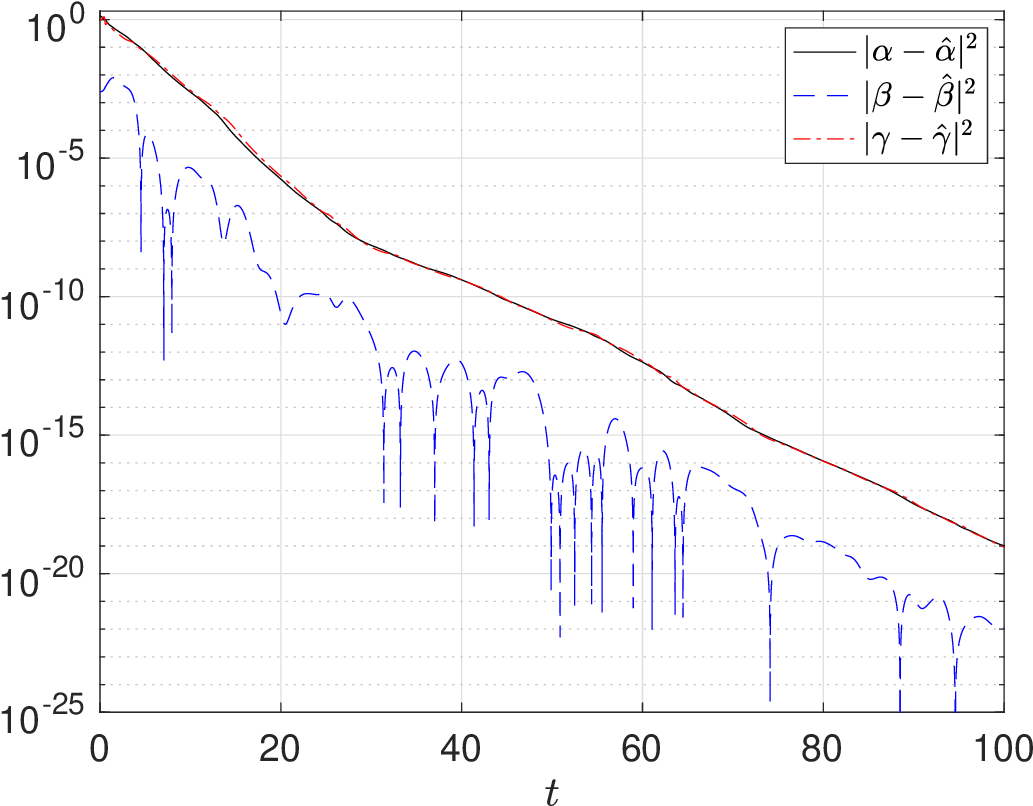}
		\caption{Parameter estimation error}
		\label{f4parerr}
	\end{subfigure}	
}
\caption{\textbf{(a)} Exponentially-fast convergence $\bar\mE^2(t) \rw 0$ of the normalized synchronization error with coupling constant $D=1$ and adaptation rate $\mu = 200$. The initial parameter values in the slave model are $\hat{\bm{\theta}}(0)=[0,0,0]^\top$. The master parameters are $\bm{\theta}=[1.15, -0.05, 0.98]^\top$. \textbf{(b)} Squared estimation error for each of the unknown parameters in $\bm{\theta}=[ \alpha,\beta,\gamma ]^\top$.}
\label{f4par}
\end{figure}

For the same simulation run as in Figure \ref{f3par}, Figure \ref{f4e2} shows that the normalized MSE, $\bar\mE^2(t)$ as given in Eq. \eqref{eqNormPower}, decreases exponentially fast with time $t$. \cblack{Recall that $\bar\mE^2(t)$ is the normalized synchronization error between the master and slave systems.} Figure \ref{f4parerr} displays the parameter estimation errors $|\alpha-\hat\alpha(t)|^2$, $|\beta-\hat\beta(t)|^2$ and $|\gamma-\hat\gamma(t)|^2$ and shows that $\hat{\bm{\theta}}(t)\rw \bm{\theta}$ exponentially fast as well. Note that, while estimation is very accurate, the errors are still decreasing after $T=100$ time units. 

\begin{figure}
\centerline{	
	\begin{subfigure}{0.45\linewidth}
		\includegraphics[width=\linewidth]{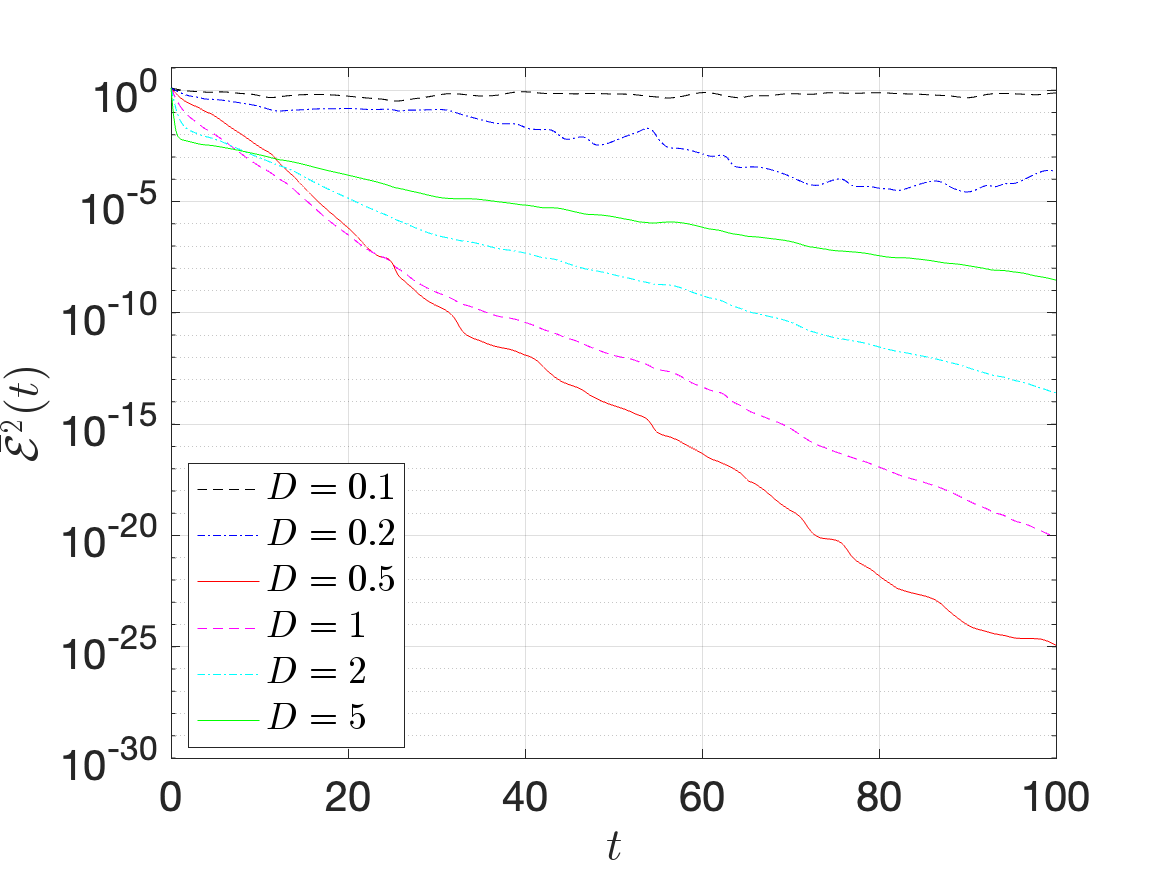}
		\caption{Normalized error $\bar{\mathcal{E}}^2(t)$ with $\mu=200$.}
		\label{f5e2D}
	\end{subfigure}
	\begin{subfigure}{0.45\linewidth}
		\includegraphics[width=\linewidth]{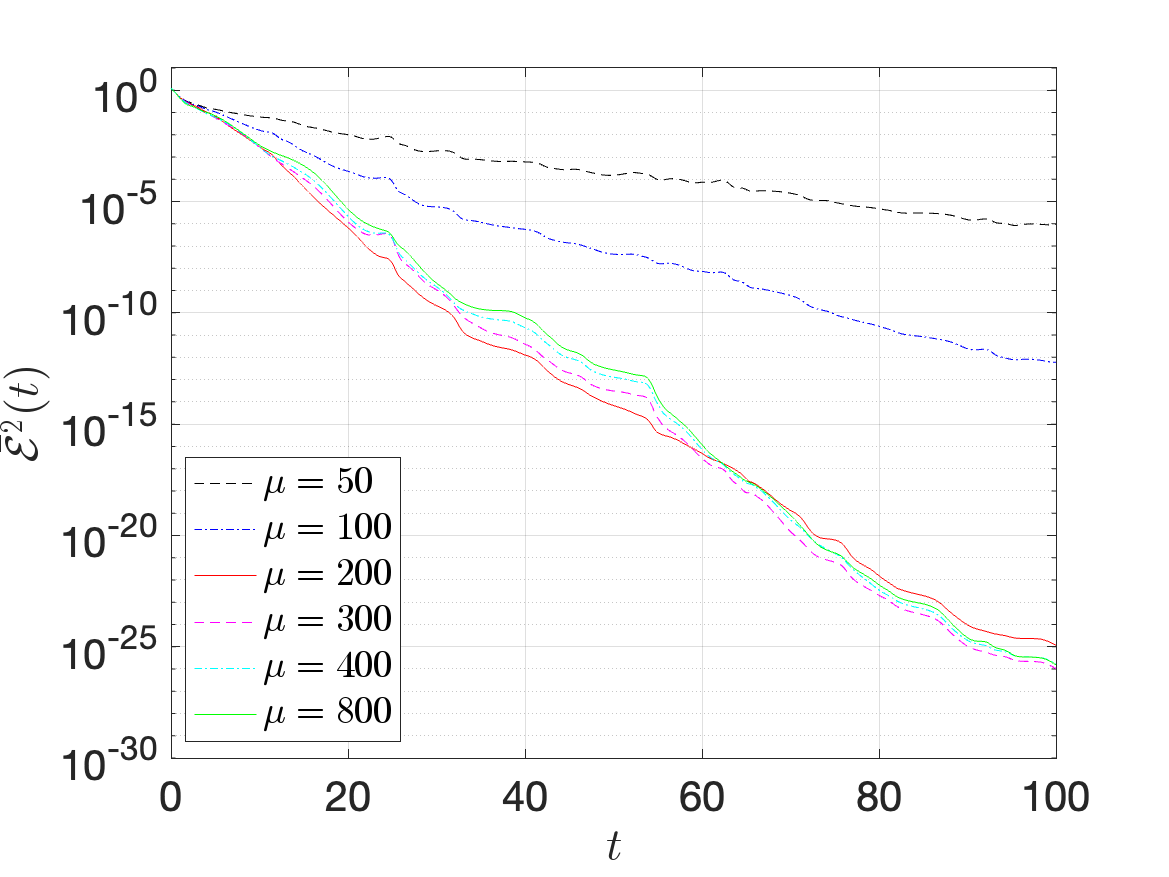}
		\caption{Normalized error $\bar{\mathcal{E}}^2(t)$ with $D=0.5$.}
		\label{f5e2mu}
	\end{subfigure}	
}
\centerline{
	\begin{subfigure}{0.45\linewidth}
		\includegraphics[width=\linewidth]{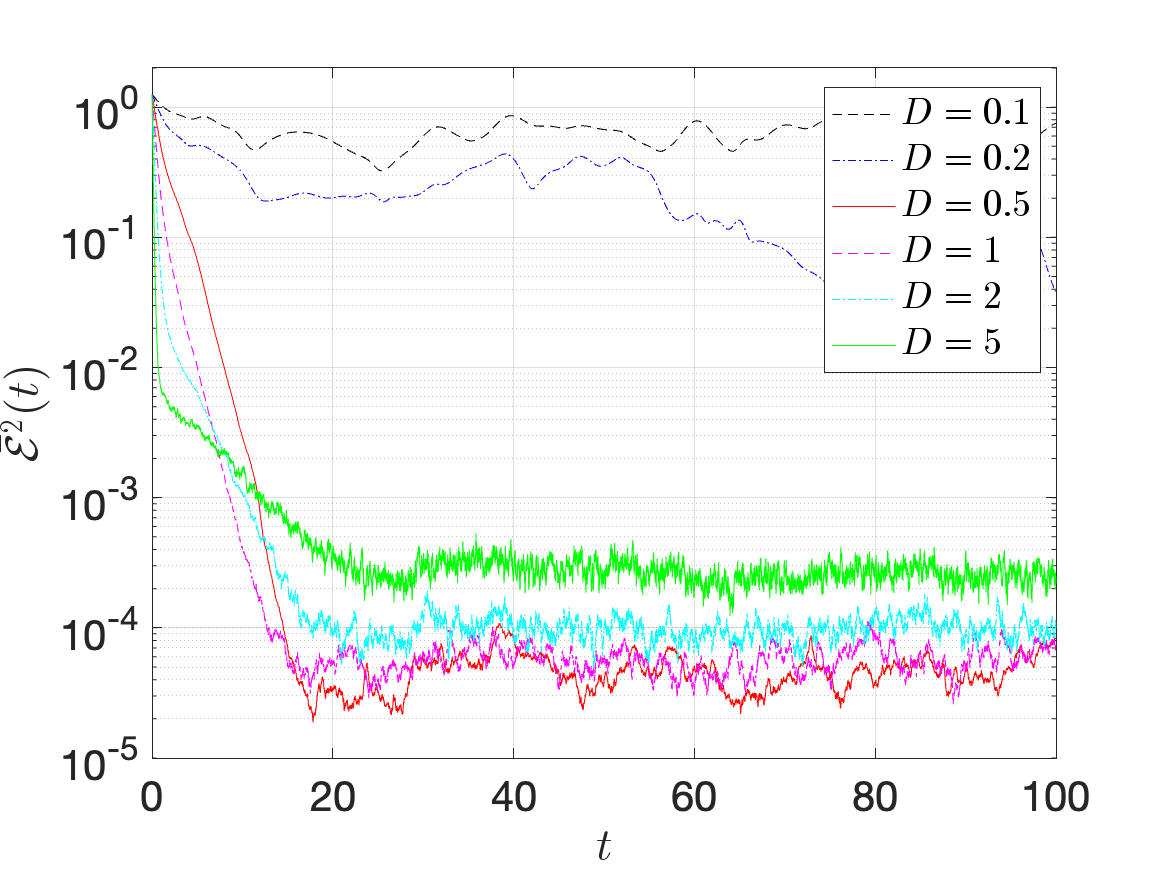}
		\caption{$\bar{\mathcal{E}}^2(t)$ with $\mu=200$ and noisy data.}
		\label{f5e2Dnoisy}
	\end{subfigure}
	\begin{subfigure}{0.45\linewidth}
		\includegraphics[width=\linewidth]{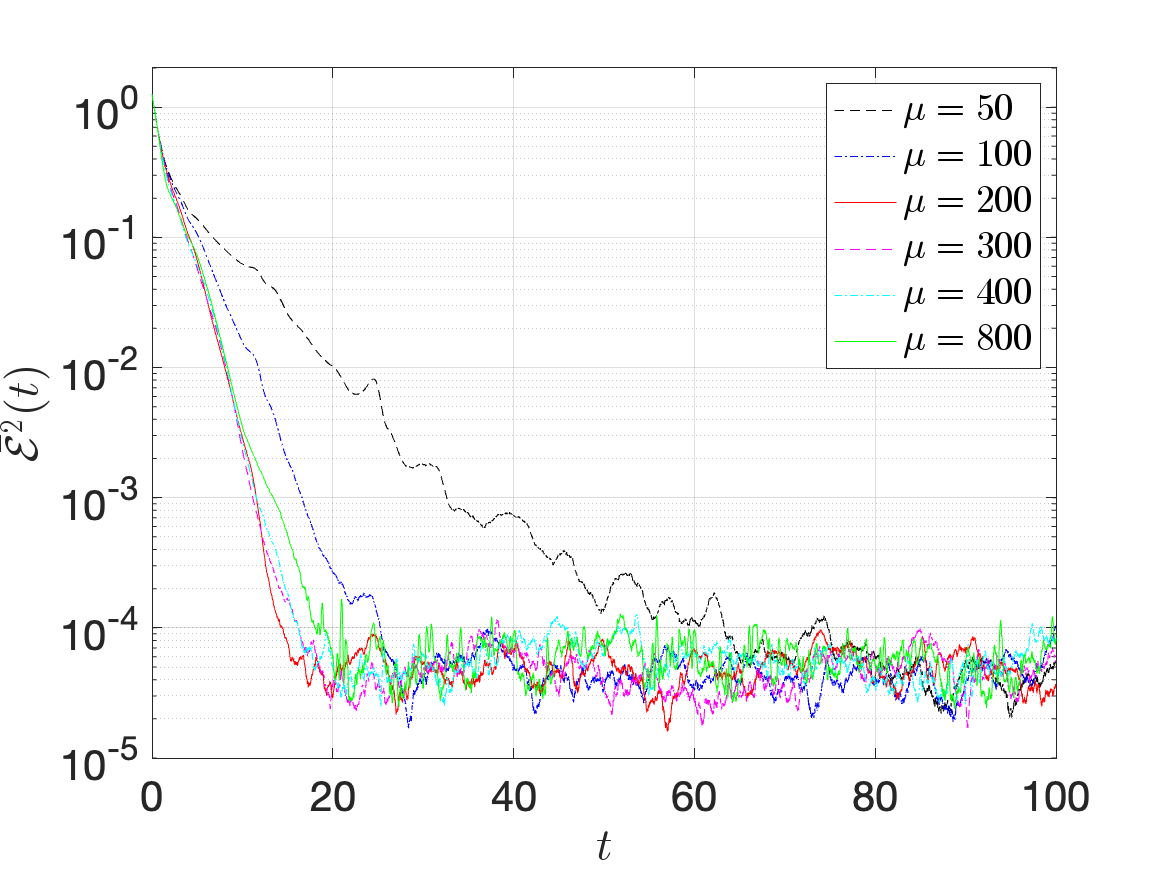}
		\caption{$\bar{\mathcal{E}}^2(t)$ with $D=0.5$ and noisy data.}
		\label{f5e2munoisy}
	\end{subfigure}

}
\caption{\textbf{(a)} Normalized synchronization error $\bar\mE^2(t)$ for different values of the coupling strength ($D$) with adaptation rate $\mu=200$. The initial parameter values in the slave model are $\hat{\bm{\theta}}(0)=[0,0,0]^\top$. \textbf{(b)} $\bar\mE^2(t)$ for different values of the adaptation rate ($\mu$) with coupling strength $D=0.5$. \textbf{(c)} $\bar\mE^2(t)$ for different values of the coupling strength ($D$) with adaptation rate $\mu=200$ and noisy data. \textbf{(d)} $\bar\mE^2(t)$ for different values of the adaptation rate ($\mu$) with coupling strength $D=0.5$ and noisy data. The average SNR in the observations of plots (c) and (d) is 12~dB.}
\label{f5par}
\end{figure}

\subsubsection{\cblack{Coupling strength and adaptation rate}}

The slave model can be `tuned' by selecting different values of the coupling coefficient $D$ and the adaptation rate $\mu$. This is illustrated in Figure \ref{f5par}. Specifically, Figure \ref{f5e2D} shows the normalized MSE, $\bar\mE^2(t)$, for several values of the coupling coefficient $D$ when the adaptation rate is kept fixed at $\mu=200$. We see that a weak coupling ($D=0.1$) does not guarantee synchronization or leads to slow convergence of the error ($D=0.2$). Best results are obtained with $D=0.5$ or $D=1$, while further increasing the coupling strength results in slower convergence (unlike the results in Section \ref{ssNumericalSynch} with known parameters). 

Next, we fix the value of the coupling coefficient to $D=0.5$ in order to study the effect of varying the adaptation rate $\mu$. Figure \ref{f5e2mu} shows the evolution of $\bar\mE^2(t)$ for several values of $\mu$. We observe that the smaller values ($\mu=50,~~100$) yield a slow decrease of the MSE. Increasing the adaptation rate to $\mu=200$ significantly improves the convergence speed, but further increments ($\mu=30,~~400,~~800$) do not result either in better accuracy or faster convergence.

\subsubsection{\cblack{Noisy observations}} \label{sssNoisy}

\cblack{For the remaining computer experiments we assume that the observed data from the master system are contaminated with Gaussian noise.}

\cblack{First, we have repeated the simulations of Figures \ref{f5e2D} and \ref{f5e2mu} with noisy data.} Specifically, we assume that the observations received at the slave model are of the form $u_K(t,x_j) + w_j(t)$, $j=0, \ldots, J-1$, where the noise terms $w_j(t)$ are independent white Gaussian stochastic processes with power spectral density $S_w(i\omega)\approx 0.17$, which corresponds to a signal-to-noise ratio (SNR) of 12~dB. Figure \ref{f5e2Dnoisy} displays the normalized synchronization error $\bar\mE^2(t)$ for several values of the coupling strength $D$ when the adaptation rate is kept fixed at $\mu=200$. We see that small coupling ($D=0.1, 0.2$) does not lead to synchronization; $D=0.5, 1$ yield similar results, with fast convergence and an error floor (due to the observation noise) below $10^{-4}$. Increasing the coupling strength further ($D=2, 5$) still yields synchronization, but the error is higher. Figure \ref{f5e2munoisy} shows the normalized error $\bar\mE^2(t)$ when $D=0.5$ is fixed and the adaptation rate parameter $\mu$ takes values between 50 and 800. In this case, we observe that a small rate ($\mu=50$) yields slow convergence of the error, although synchronization is attained after approximately 70 time units. For $\mu \ge 50$, the performance is similar for all values. Slightly faster convergence to the synchronized state is obtained for $\mu=200,300, 400$, but the error floor is very similar for $\mu=100, 800$ as well. 

\begin{figure}
\centerline{	
	\begin{subfigure}[t]{0.32\linewidth}
		\includegraphics[width=\linewidth]{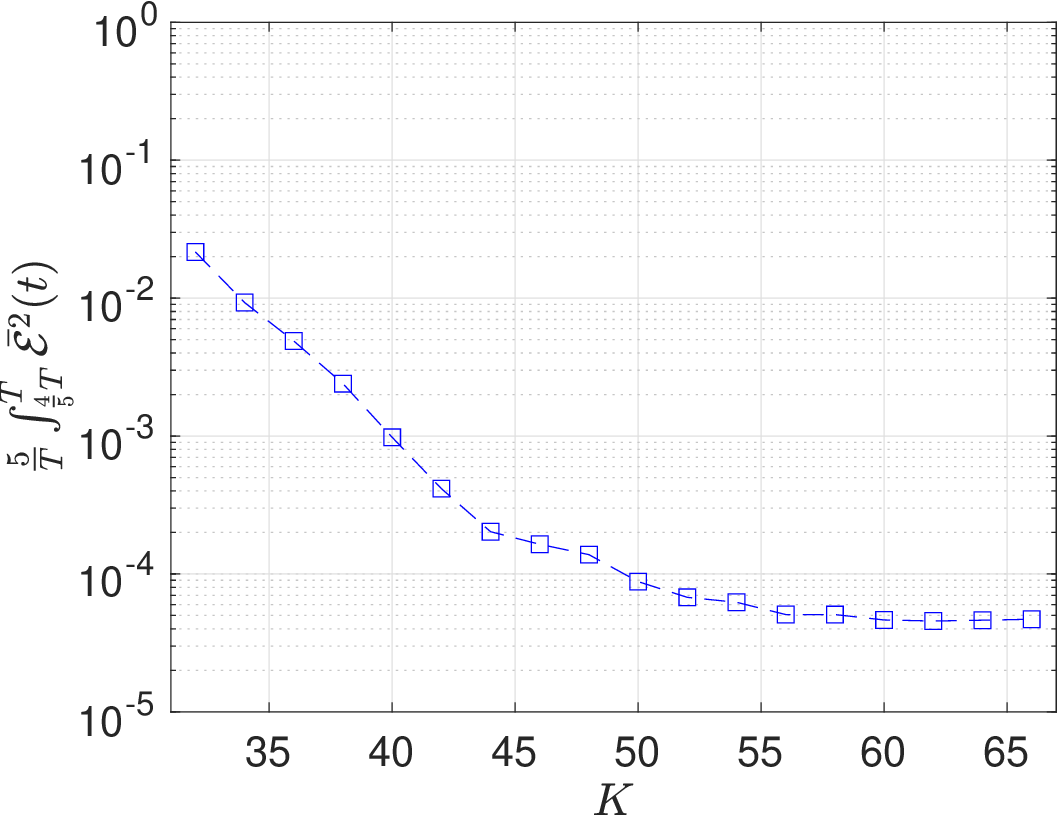}
		\caption{Averaged normalized error for $M=64$ and $K=32, \ldots, 66$. Noisy data.}
		\label{f6eK}
	\end{subfigure}
	\begin{subfigure}[t]{0.33\linewidth}
		\includegraphics[width=\linewidth]{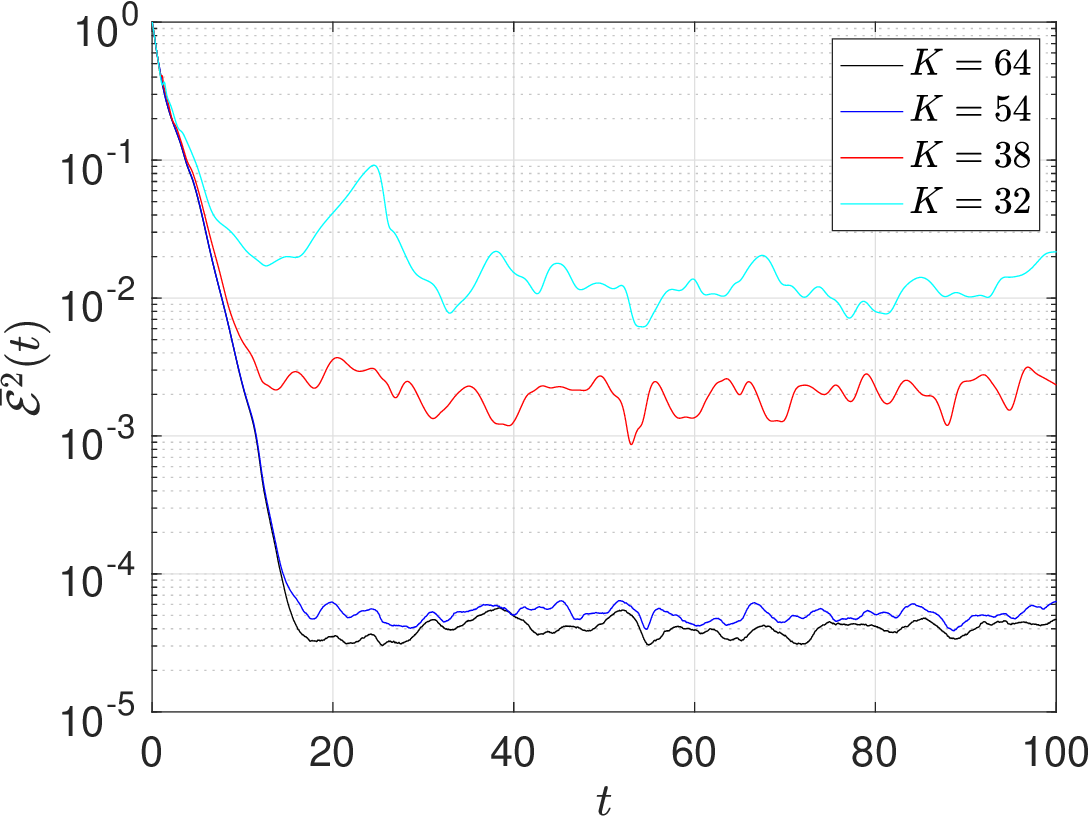}
		\caption{$\bar{\mathcal{E}}^2(t)$ vs. time for $M=64$ and $K=32, 38, 54, 64$. Noisy data.}
		\label{f6e2K}
	\end{subfigure}	
}
\centerline{	
	\begin{subfigure}[t]{0.32\linewidth}
		\includegraphics[width=\linewidth]{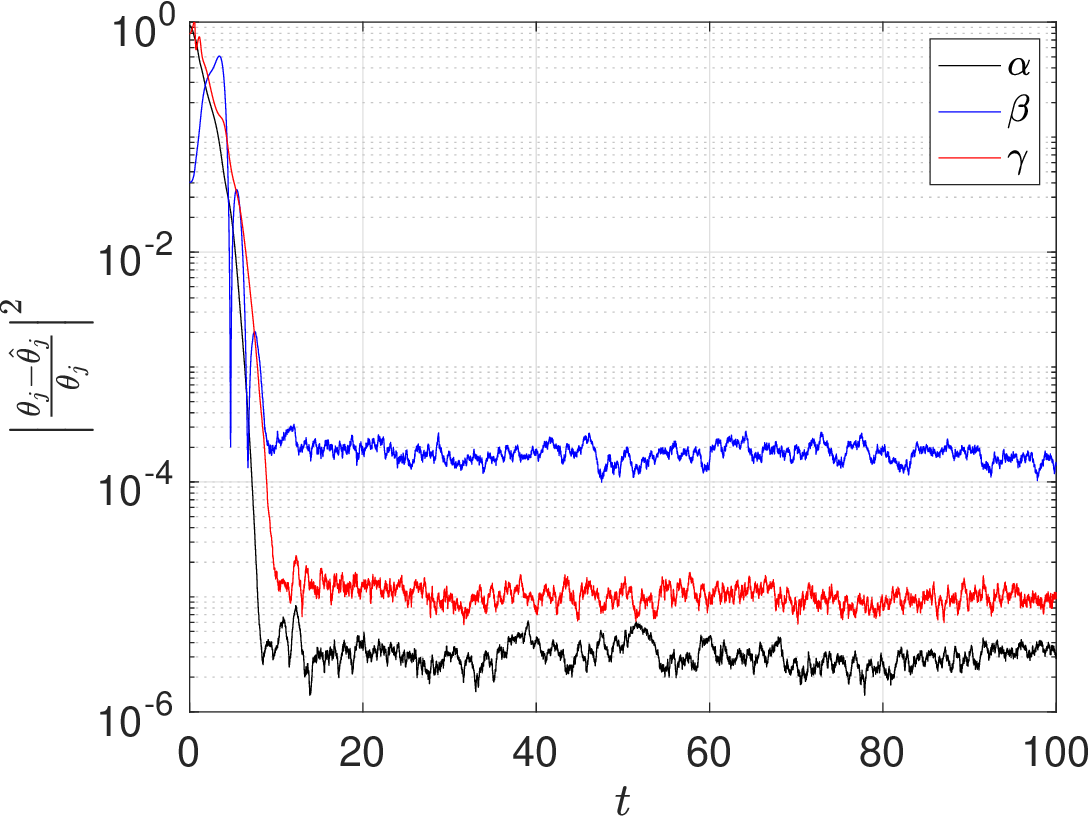}
		\caption{Parameter estimation error with $M=64$ and $K=64$. Noisy data.}
		\label{f6parerr64}
	\end{subfigure}
	\begin{subfigure}[t]{0.32\linewidth}
		\includegraphics[width=\linewidth]{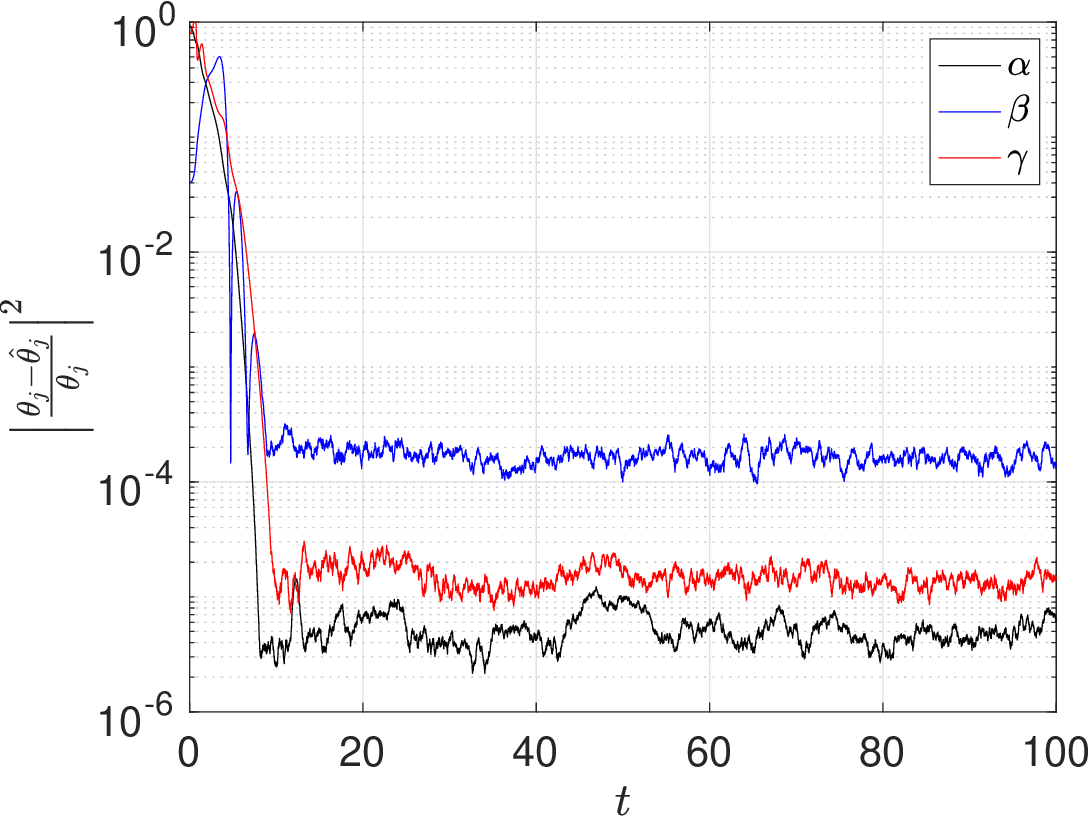}
		\caption{Parameter estimation error with $M=64$ and $K=54$. Noisy data.}
		\label{f6parerr54}
	\end{subfigure}	
	\begin{subfigure}[t]{0.32\linewidth}
		\includegraphics[width=\linewidth]{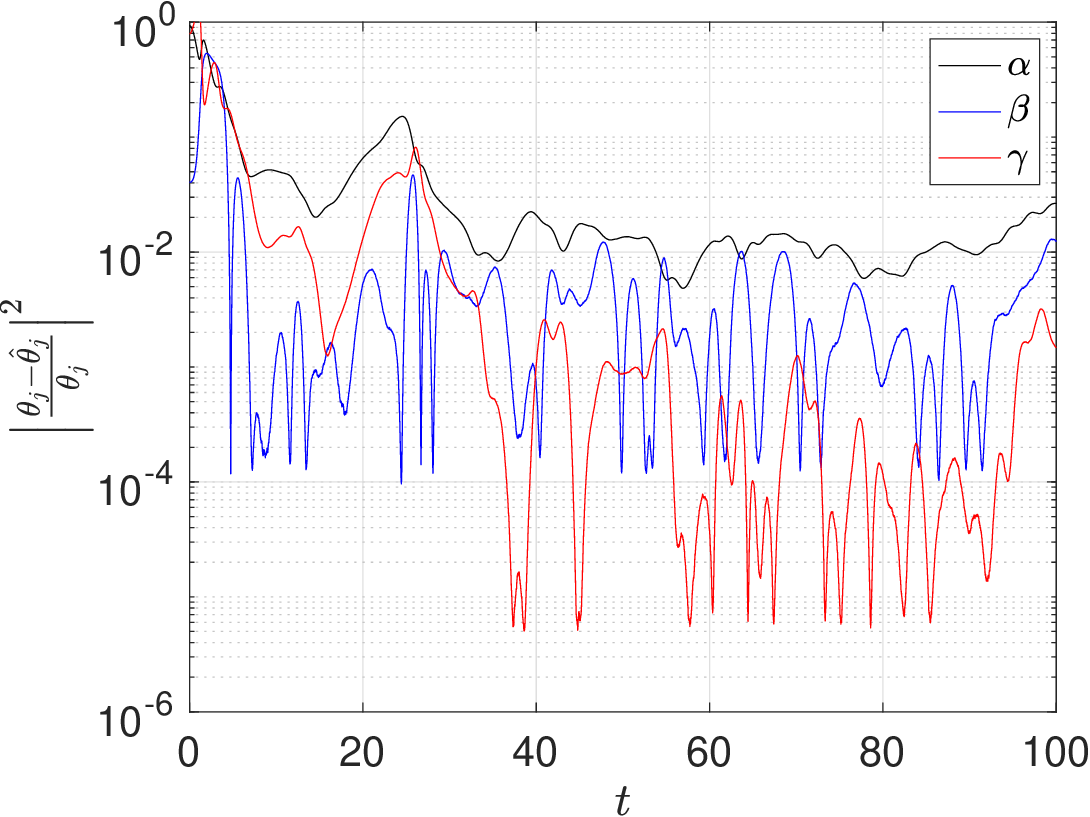}
		\caption{Parameter estimation error with $M=64$ and $K=32$. Noisy data.}
		\label{f6parerr32}
	\end{subfigure}	
}
\caption{\textbf{(a)} Average over the interval $t \in \left[\frac{4T}{5}, T\right]$ of the normalized MSE, $\frac{5}{T}\int_{\frac{4}{5}T}^T \bar\mE^2(t)\md t$, for $K=32, \ldots, 64$ in the slave model. The master model is implemented with $M=64$. The coupling strength is $D=0.5$ and the adaptation rate is $\mu=200$.  The initial parameter values in the slave model are $\hat{\bm{\theta}}(0)=[0,0,0]^\top$ and the master parameters are $\bm{\theta}=[1.15, -0.05, 0.98]^\top$. The observations are contaminated with Gaussian noise (the average SNR is 12~dB). The error is almost equally small for all $K \ge 54$. \textbf{(b)} $\bar\mE^2(t)$ for $K=32, 38, 54, 64$ for the same set of simulations. \textbf{(c)} Normalized quadratic parameter estimation error for the three model parameters, $\bm{\theta}=[\alpha,\beta,\gamma]^\top$, when $K=64$. \textbf{(d)} Normalized quadratic parameter estimation error when $K = 54$. \textbf{(e)} Normalized quadratic parameter estimation error when $K = 32$.}
\label{f6par}
\end{figure}

\subsubsection{\cblack{Number of Fourier modes}} \label{sssModes}

Figure \ref{f6par} shows the results of a set of computer experiments designed to study the effect of underestimating the number of coefficients $K$ needed in the slave model. To this end, we generate the signal in the master system $u_M(t,x)$ using, again, $M=64$ Fourier modes and then run slave models with order $K = 32, 34, \ldots, 66$. The coupling strength and the adaptation rate in all the slave models are $D=0.5$ and $\mu=200$, respectively. For all the simulations in this figure we have assumed that the observations are perturbed with white Gaussian noise with power spectral density $S_w(i\omega)\approx 0.17$ (average signal-to-noise ratio 12~dB). The results are averaged over 100 simulation runs, where the noise in the observations is generated independently for each run (while all other simulation parameters are kept the same).  

Figure \ref{f6eK} shows the synchronization error versus the order $K$ of the slave model. The error is an average over the interval $\left[\frac{4}{5}T,T\right]$ of the normalized MSE $\bar\mE^2(t)$, i.e., $\bar \mE^2 := \frac{5}{T}\int_{\frac{4T}{5}}^T \bar \mE^2(t) \md t$. We observe how the synchronization error decreases consistently from $K=32$ up to $K \approx 54$ and then stays approximately flat. This indicates that, even if the master model has Fourier modes up to order $M=64$, $K \approx 54$ modes are sufficient to design a slave model that attains identical synchronization and yields accurate parameter estimates \cblack{--this is coherent with early results on the existence and computation of inertial manifolds for the KS equation \cite{Foias88}.} In particular, Figure \ref{f6e2K} shows the evolution of the error $\bar\mE^2(t)$ versus time for $K=32, 38, 54, 64$. We observe that the synchronization error is nearly the same for $K=54$ and $K=64$, and two orders of magnitude smaller than the error for $K=32$. We point out, nevertheless, that $\bar\mE^2(t)$ is normalized hence, even for $K=32$, the power of the actual error is just $\approx 1\%$ of the power of the master signal $u_M(t,x)$. 

Figures \ref{f6parerr64}, \ref{f6parerr54} and \ref{f6parerr32} show the normalized squared estimation error for the parameters in $\bm{\theta}=\left[\alpha,\beta,\gamma\right]^\top$ when $K=64$ (Fig. \ref{f6parerr64}), $K=54$ (Fig. \ref{f6parerr54}) and $K=32$ (Fig. \ref{f6parerr32}), respectively. Again, we see that the estimation error is almost the same with $K=64$ and $K=54$, while it is significantly higher for $K=32$. Nevertheless, we again point out that these are normalized errors and, even for $K=32$, they are around or below $1\%$ for most of the simulation interval.

\subsubsection{\cblack{Comparison with other methods}}

\cblack{Most methods for parameter estimation in the KS equation are offline \cite{Hurst22,Huttunen18,Lu17,MartinaPerez21,Rudy19,Hu08}. They are designed for a different setup (possibly with scarce observations) and have a much higher computational cost than the online scheme introduced in Section \ref{ssSlaveEstimator}. Comparisons should be carried out with other online methods.}

\cblack{The most similar scheme is the one by Pachev {\em et al.} \cite{Pachev22}. The computational cost of both Pachev {\em et al.}'s method and the proposed scheme is dominated by the computation of the convolution $\tilde{b}_k = \sum_{\ell=-K}^K b_\ell b_{\ell-k}$ (see Eq. \eqref{eqFourierSlave}), which has complexity $\mathcal{O}(K \log K)$. However, Pachev {\em et al.}'s method requires the computation of time derivatives of the master signal (i.e., either $\bm{u}_t(t)$ or $\hat{\bm{a}}_t(t)$), which are hard to compute in the presence of observational noise. Equations \eqref{eqParUpdate} and \eqref{eqCoeffUpdate} involve $\hat{\bm{a}}$ alone (and {\em not} its derivative $\hat{\bm{a}}_t$) which makes the proposed method directly applicable (and robust, as shown in Sections \ref{sssNoisy} and \ref{sssModes}) with noisy observations.}

\cblack{Within the class of statistical methods, a comparison can be carried out with the unscented Bucy-Kalman filter (UBKF) \cite{Sarkka07}. This is an online method that can be used to approximate the conditional mean and covariance matrix of the $(M+4) \times 1$ extended state vector 
$
\bm{s}(t) = \left[ 
	\begin{array}{c}
	\bm{\theta}\\
	\bar{\bm{a}}(t)\\
	\end{array}
\right]
$
at any time $t$, given the observations $\bm{u}(\tau)$, $0 < \tau \le t$. The UBKF relies on a quadrature or cubature scheme to approximate the nonlinearity of the KS equation. While other possibilities exist \cite{Jia13,Menegaz15}, we have employed a spherical-radial cubature rule of degree 3 \cite{Jia13} which uses $L=2(M+4)$ reference points. These $L$ points are deterministically computed and they have to be propagated across the KS equation (in the same vein as one would do with a Monte Carlo method) for each step of the numerical integration scheme applied to the system of ODEs in \eqref{eqFourierM}.
}

\cblack{The UBKF is one of the simplest statistical methods that can be applied to the parameter estimation problem for the KS equation, and yet its computational cost is much higher than the synchronization-based technique of Section \ref{ssSlaveEstimator} or the method by Pachev {\em  et al.} \cite{Pachev22}. To be specific, both the UBKF and the proposed method have $\mathcal{O}(T)$ complexity (they are both online), however the number of computations at each step of the Euler scheme is $\mathcal{O}(K \log K)$ for the proposed method versus $\mathcal{O}(K^2 \log K)$ for the UBKF (assuming $M=K$ Fourier modes in both cases). Also, the processing of the a $J \times 1$ observation vector $\bm{u}(t)$ takes $\mathcal{O}(J)$ operations for the proposed method, versus $\mathcal{O}(J^3)$ for the UBKF.}

\cblack{We have compared numerically the performance of the synchronization-based method given by Eqs. \eqref{eqParUpdate}-\eqref{eqCoeffUpdate} with the UBKF algorithm within the same setup of Section \ref{sssNoisy}. The UBKF is implemented as described above, with $M=64$ and $L=2(M+4)=268$ reference points. We assume $K=M=64$ in the slave model for the synchronization-based method.}

\begin{figure}
\centerline{	
	\begin{subfigure}[t]{0.4\linewidth}
		\includegraphics[width=\linewidth]{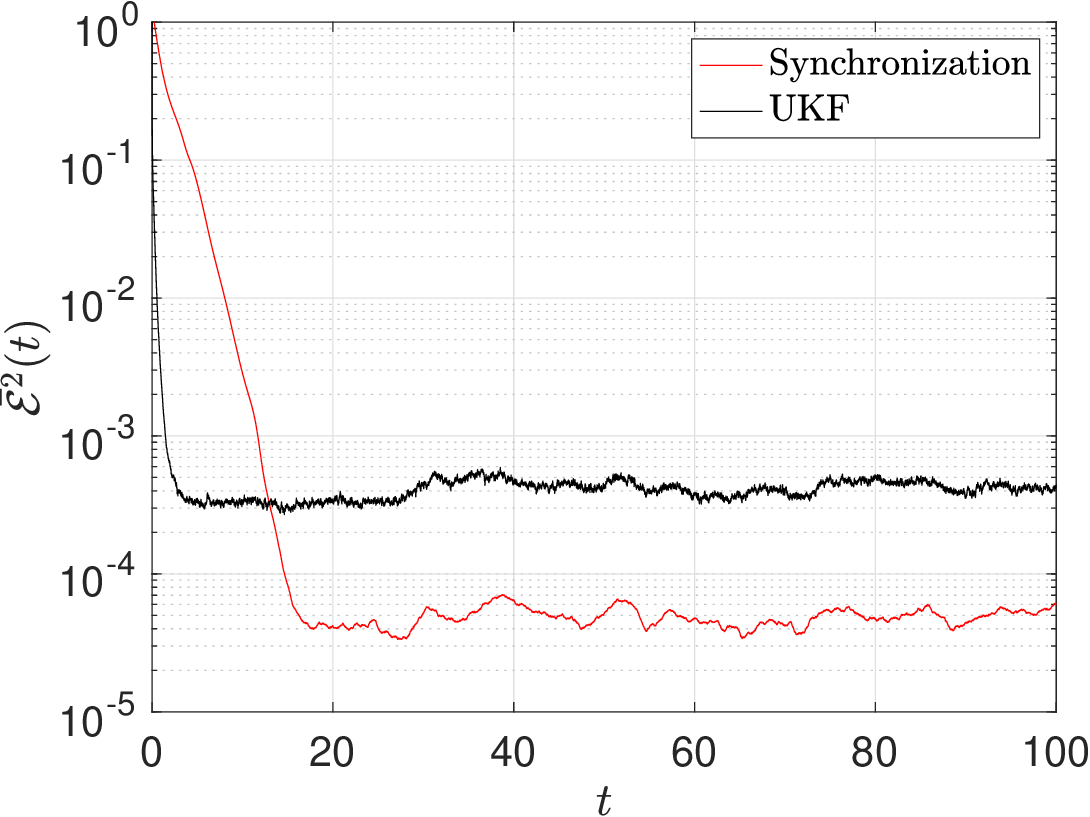}
		\caption{$\bar{\mathcal{E}}^2(t)$ vs. time for the UBKF and synchronization-based method.}
		\label{f7synch}
	\end{subfigure}
	\begin{subfigure}[t]{0.4\linewidth}
		\includegraphics[width=\linewidth]{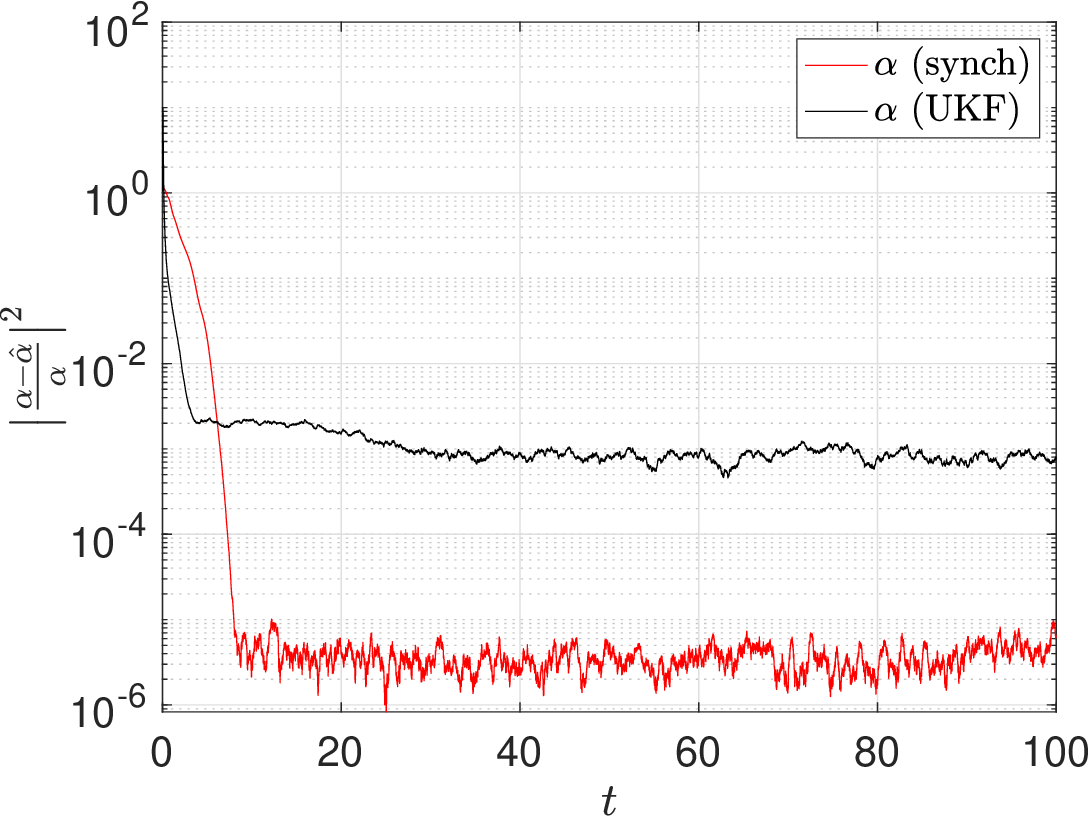}
		\caption{Normalized parameter estimation error vs. time. Parameter $\alpha$}
		\label{f7alpha}
	\end{subfigure}	
}
\centerline{	
	\begin{subfigure}[t]{0.4\linewidth}
		\includegraphics[width=\linewidth]{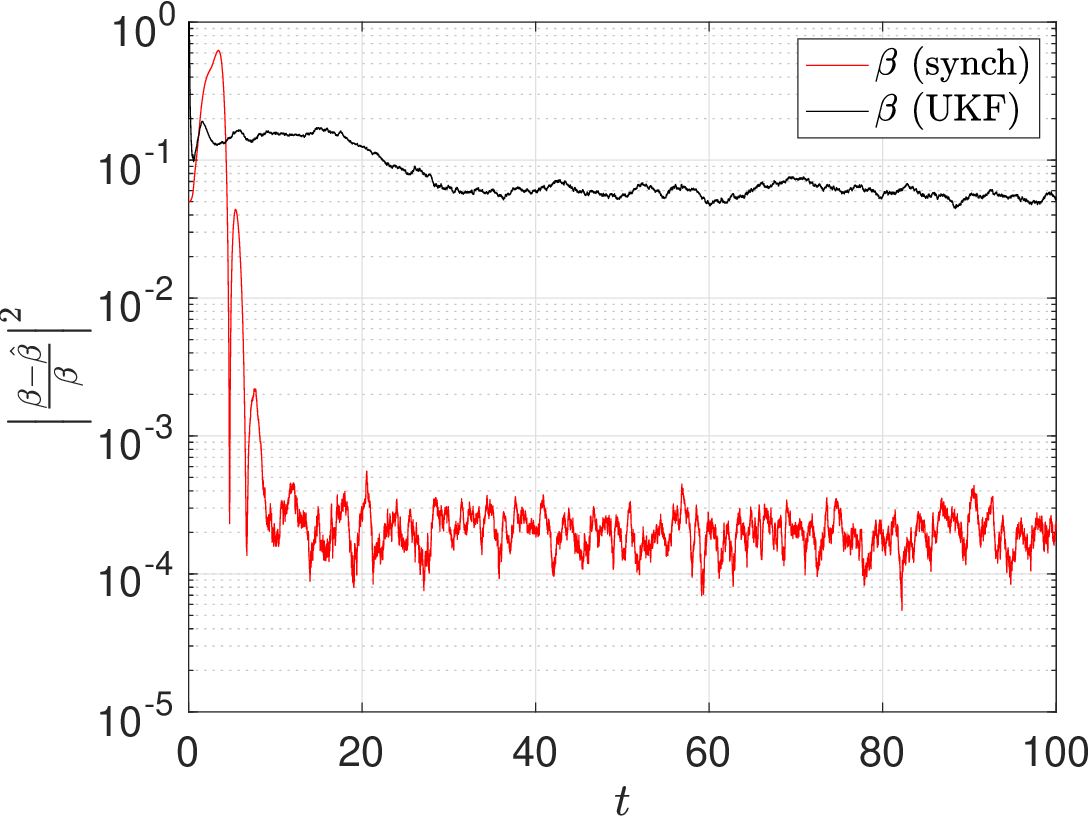}
		\caption{Normalized parameter estimation error vs. time. Parameter $\beta$}
		\label{f7beta}
	\end{subfigure}
	\begin{subfigure}[t]{0.4\linewidth}
		\includegraphics[width=\linewidth]{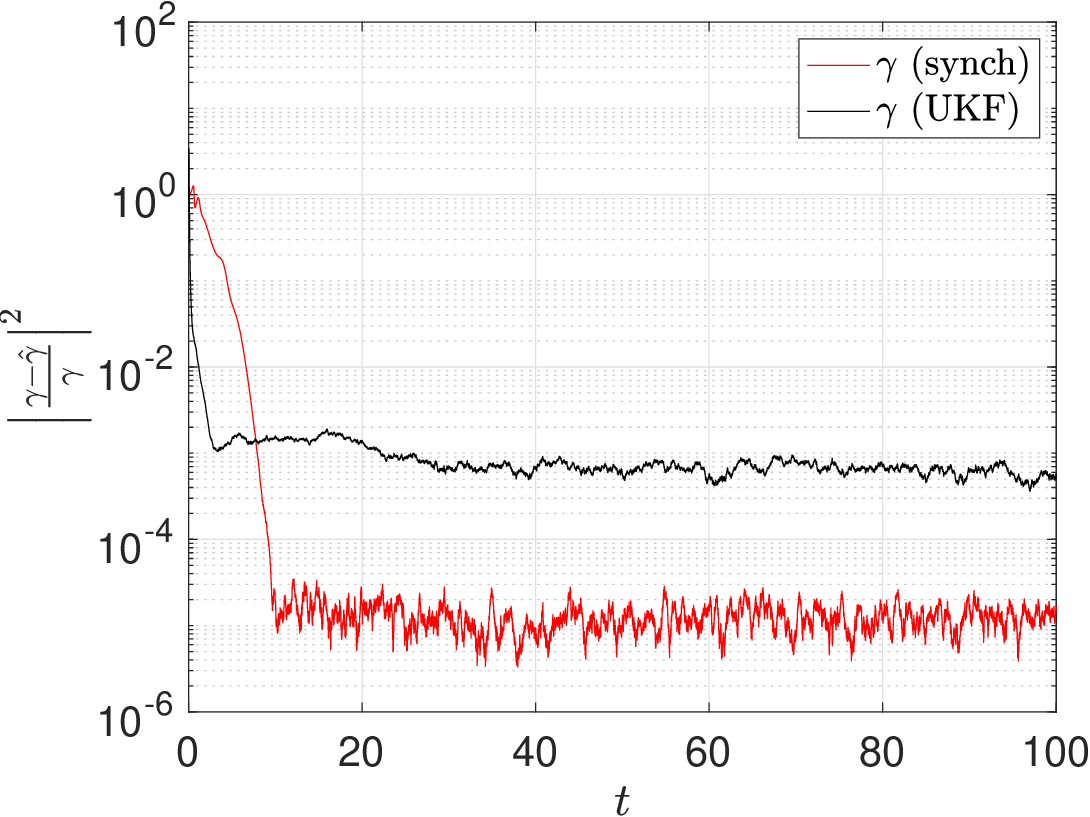}
		\caption{Normalized parameter estimation error vs. time. Parameter $\gamma$}
		\label{f7gamma}
	\end{subfigure}		
}
\caption{\cblack{\textbf{(a)} Normalized MSE $\bar{\mathcal{E}}^2(t)$ versus time for the UBKF with $2(K+4)$ sigma-points and the proposed synchronization-based method with $M=K=64$ Fourier modes. \textbf{(b, c, d)} Normalized parameter estimation errors versus. Observations are noisy, with average signal-to-noise ratio 12~dB. The coupling strength and adaptation rate are $D=0.5$ and $\mu=200$, respectively. The true parameters are $\bm{\theta}=[1.15, -0.05, 0.98]$, the initial values in the UBKF are $\hat{\bm{\theta}}(0)=[0.05,0.05,0.05]^\top$ and in the slave model are $\hat{\bm{\theta}}(0)=[0,0,0]^\top$. The results are averaged over 20 independent simulations.}}
\label{f7ukf}
\end{figure}

\cblack{
Some results of the comparison are displayed in Figure \ref{f7ukf}. Specifically, Figure \ref{f7synch} shows the normalized synchronization error $\bar{\mathcal{E}}^2(t)$ attained by the proposed method and the UBKF. We observe that the UBKF converges more quickly, however the steady-state error of the proposed technique is one order of magnitude smaller. Both curves have been averaged over 20 independent simulation runs. With our code (running with Matlab R2023b on a MacBook Pro laptop with 64GB of memory and Apple M2 Max processor), the average run time of the synchronization-based method is $\approx 3$~s for $T=100$ and $h=0.005$, versus $\approx 300$~s for the UBKF in the same setup (this can be reduced by parallelization, though).
}

\cblack{For the same set of simulations, Figures \ref{f7alpha}, \ref{f7beta} and \ref{f7gamma} show the normalized parameter estimation error for $\alpha$, $\beta$ and $\gamma$, respectively. Again, we see that convergence is faster for the UBKF but the synchronization-based method is more accurate, with errors at least two orders of magnitude smaller for all three parameters.}

\cblack{
\begin{remark}
We have compared the signal and parameter estimates directly in Figure \ref{f7ukf}. It should be noted that the UBKF also yields a covariance matrix for these estimates.
\end{remark}
}

\subsubsection{\cblack{Dynamical regimes}}

\begin{figure}
\centerline{	
	\begin{subfigure}[t]{0.32\linewidth}
		\includegraphics[width=\linewidth]{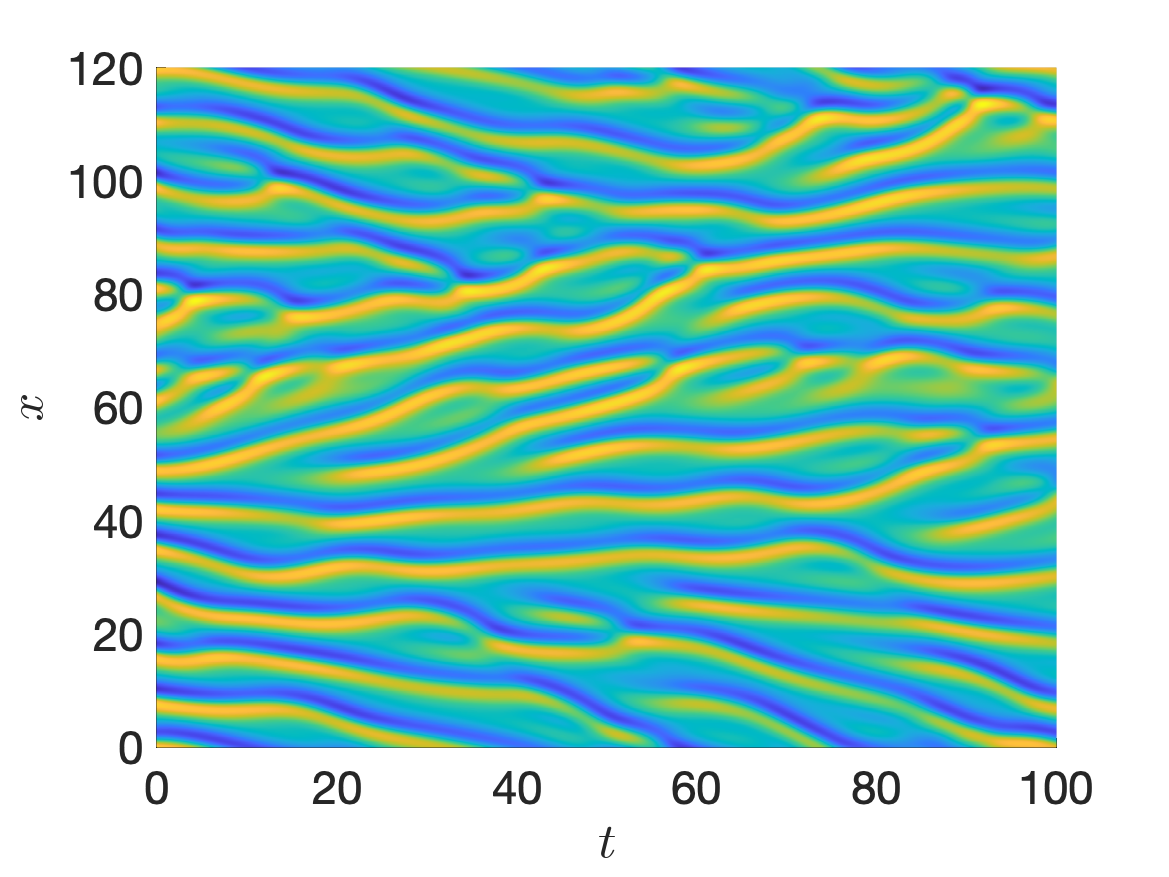}
		\caption{Master signal with $\beta=0$.}
		\label{f8a}
	\end{subfigure}
	\begin{subfigure}[t]{0.32\linewidth}
		\includegraphics[width=\linewidth]{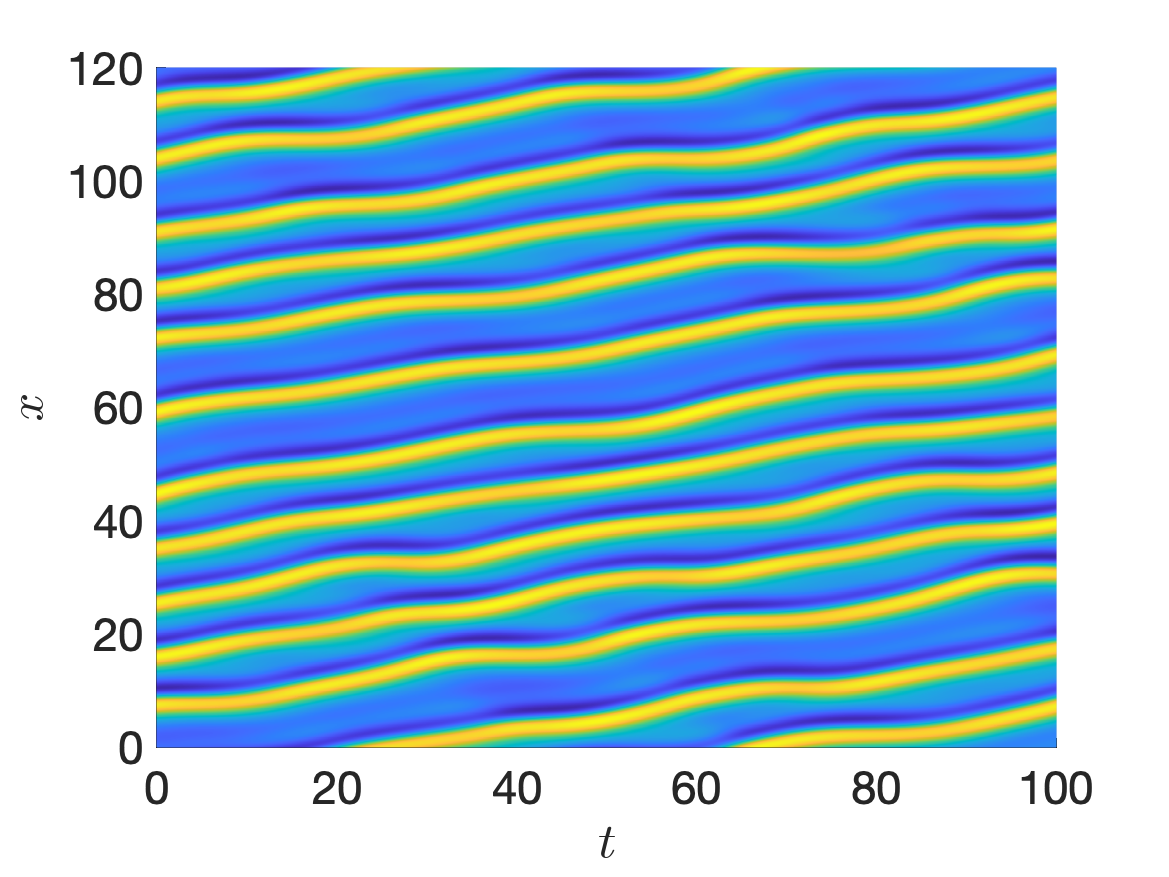}
		\caption{Master signal with $\beta=0.2$.}
		\label{f8b}
	\end{subfigure}	
\begin{subfigure}[t]{0.32\linewidth}
		\includegraphics[width=\linewidth]{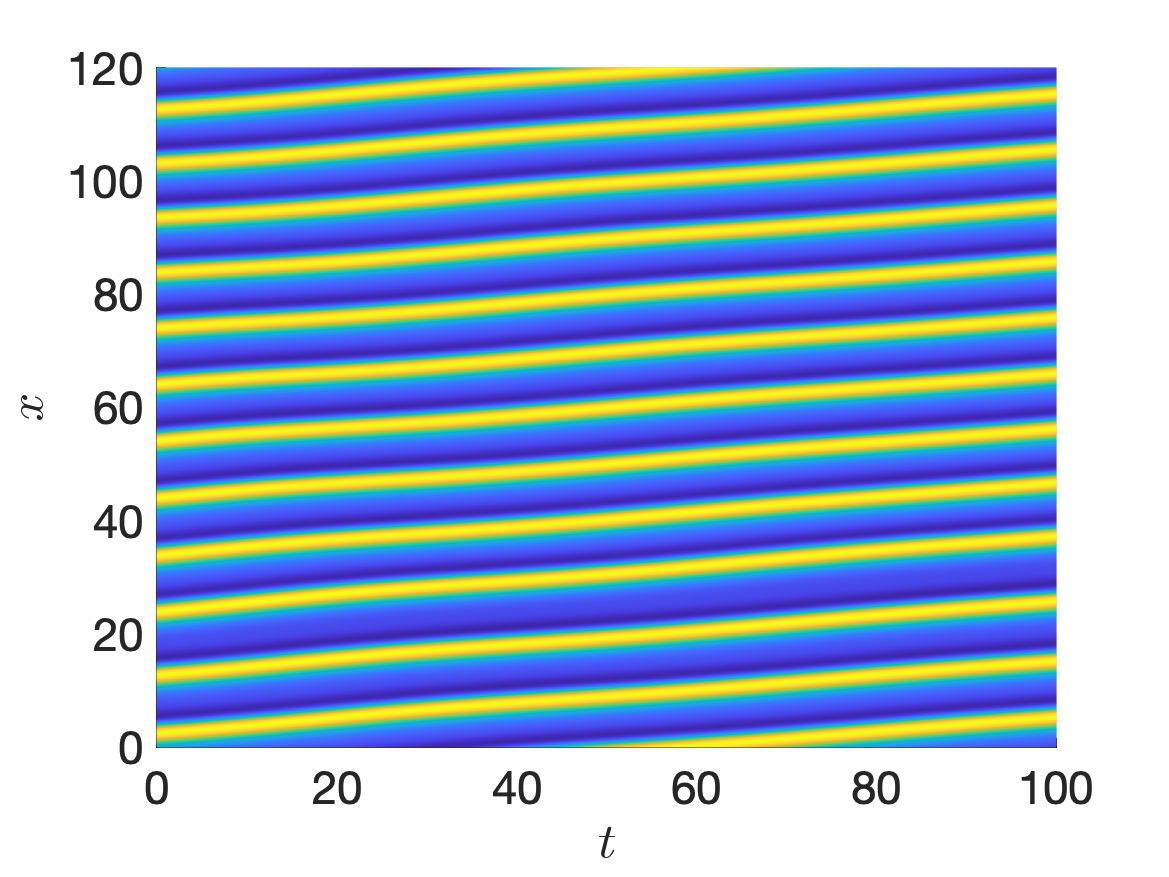}
		\caption{Master signal with $\beta=0.8$.}
		\label{f8c}
	\end{subfigure}	
}
\centerline{	
	\begin{subfigure}[t]{0.3\linewidth}
		\includegraphics[width=\linewidth]{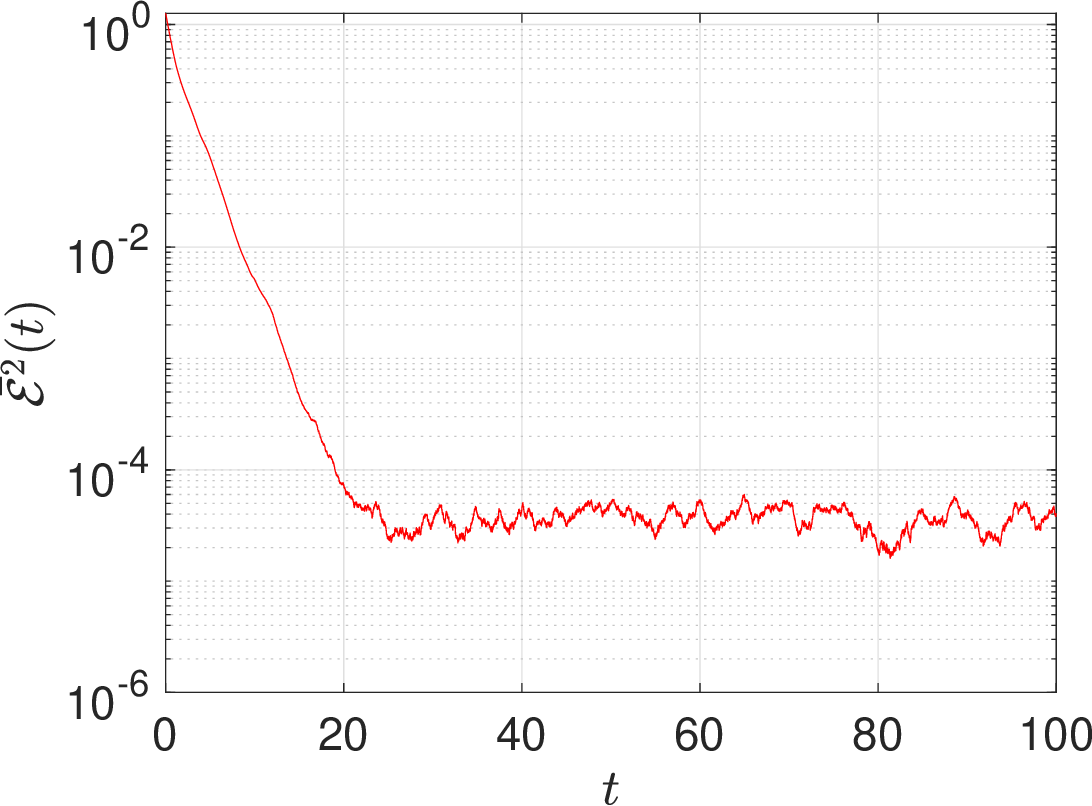}
		\caption{Synchronization error with $\beta=0$.}
		\label{f8d}
	\end{subfigure}
	\begin{subfigure}[t]{0.3\linewidth}
		\includegraphics[width=\linewidth]{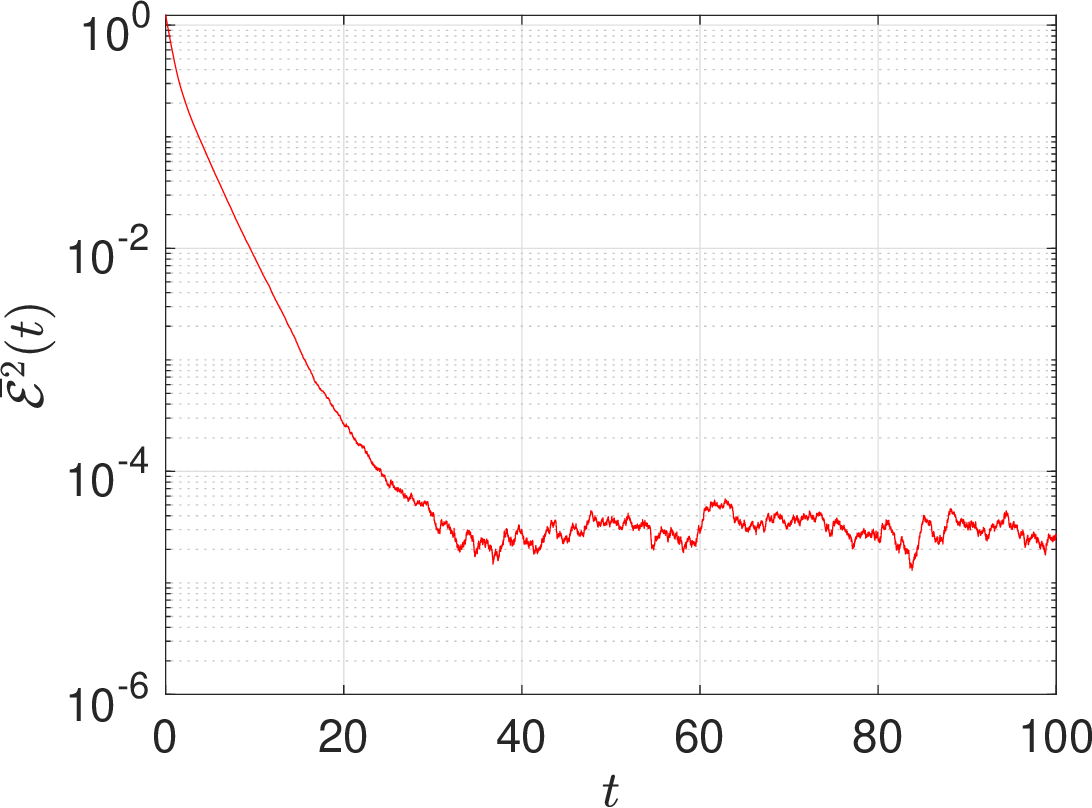}
		\caption{Synchronization error with $\beta=0.2$.}
		\label{f8e}
	\end{subfigure}	
\begin{subfigure}[t]{0.3\linewidth}
		\includegraphics[width=\linewidth]{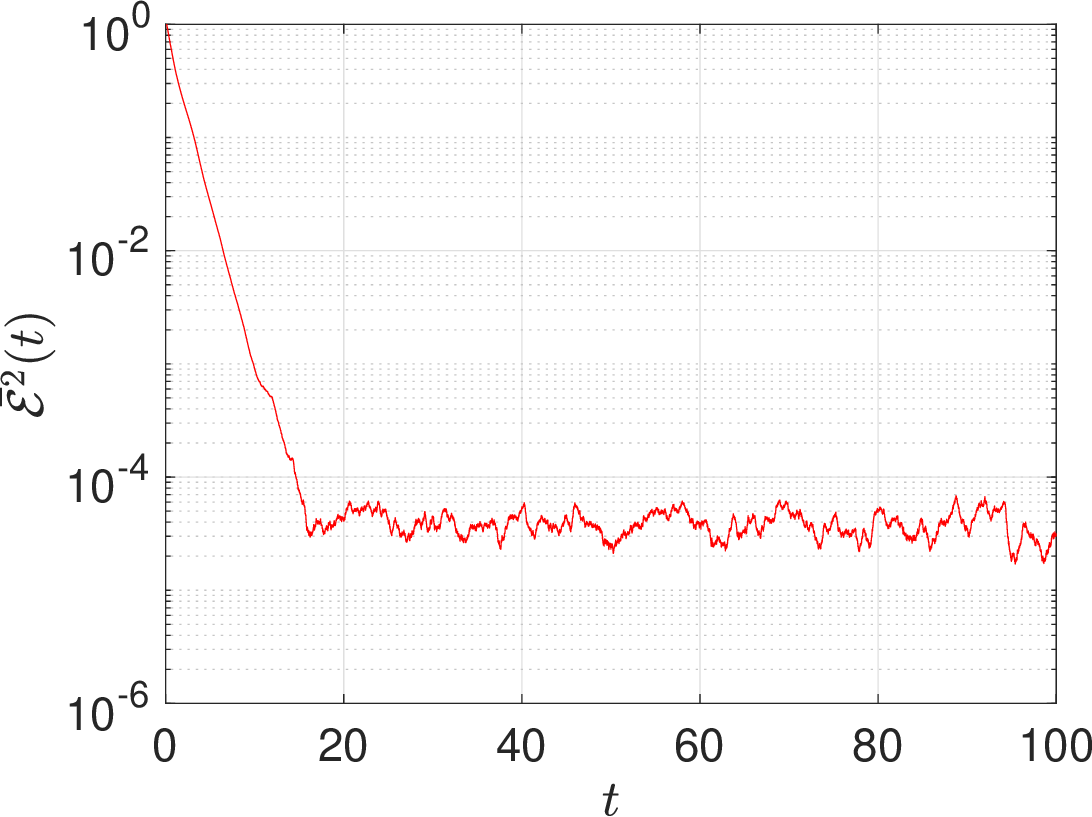}
		\caption{Synchronization error with $\beta=0.8$.}
		\label{f8f}
	\end{subfigure}	
}
\centerline{	
	\begin{subfigure}[t]{0.32\linewidth}
		\includegraphics[width=\linewidth]{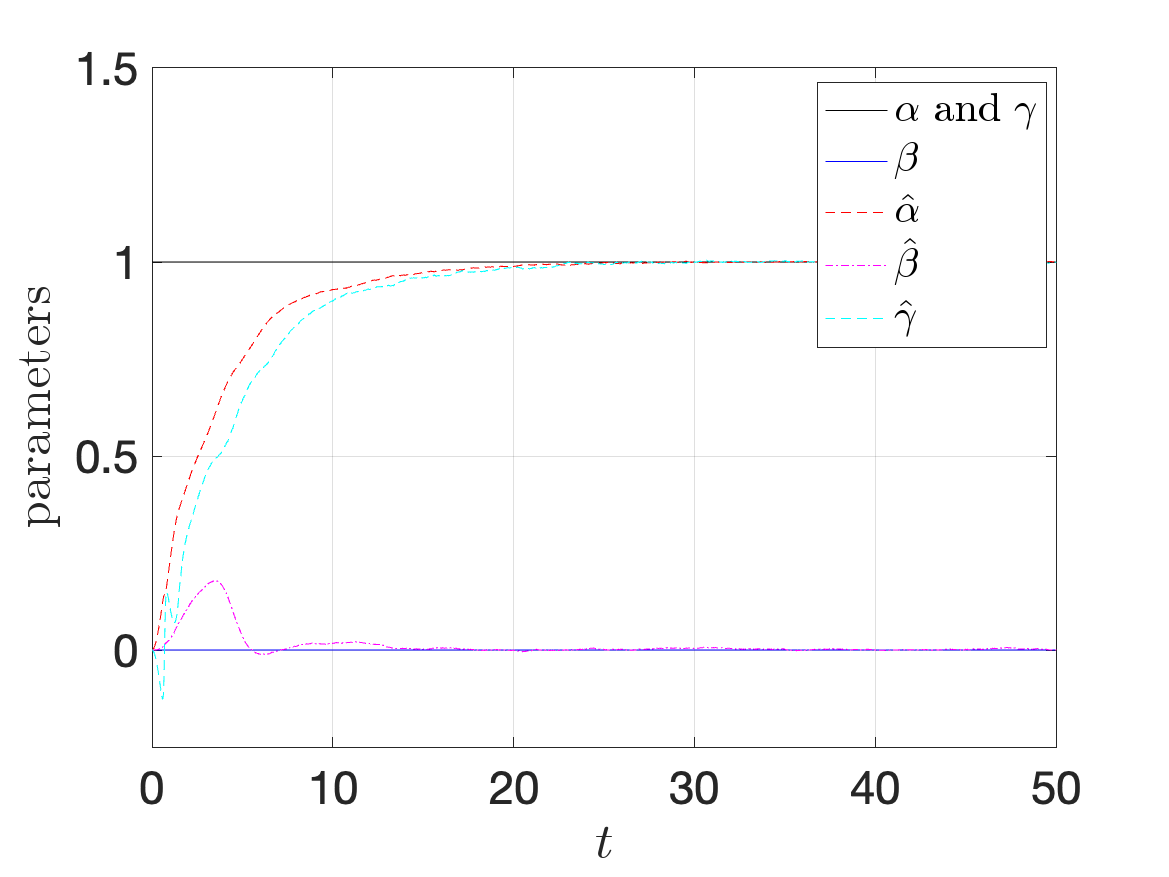}
		\caption{Parameter errors with $\beta=0$.}
		\label{f8g}
	\end{subfigure}
	\begin{subfigure}[t]{0.32\linewidth}
		\includegraphics[width=\linewidth]{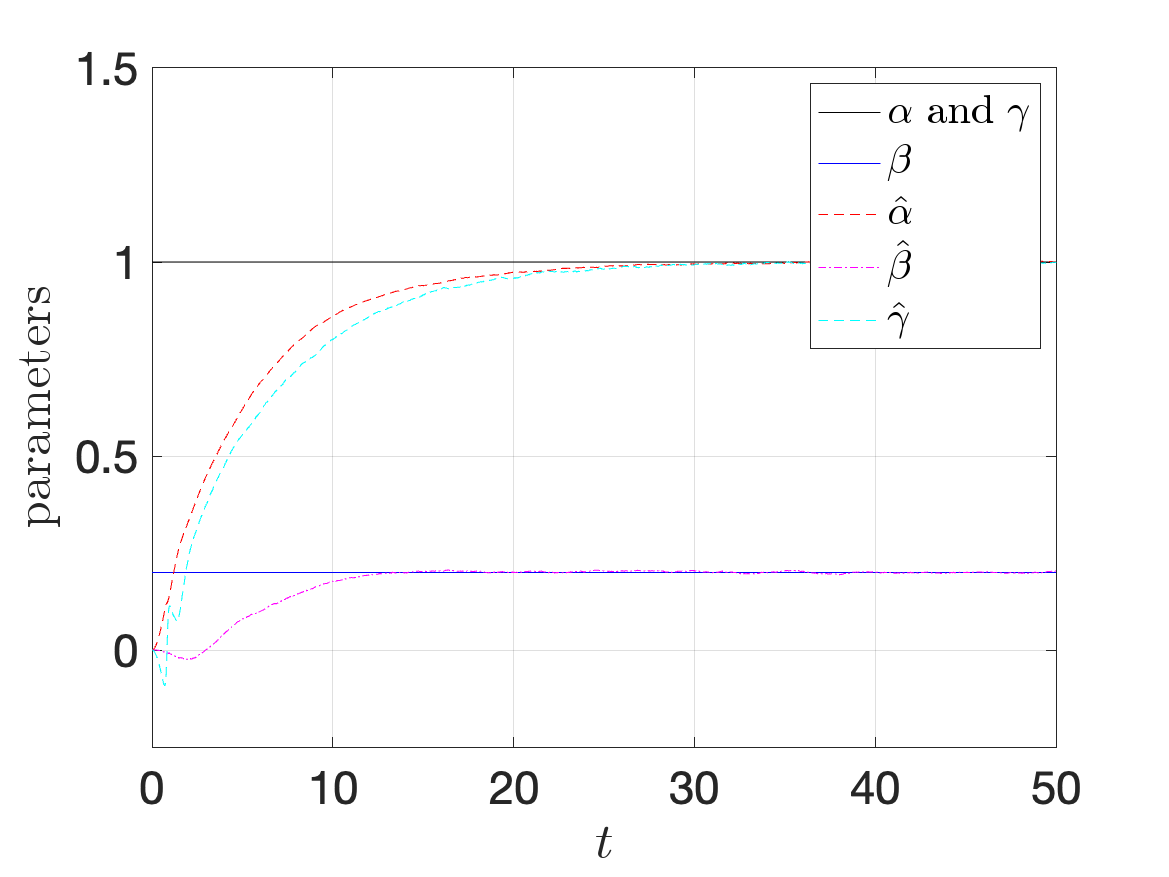}
		\caption{Parameter errors with $\beta=0.2$.}
		\label{f8h}
	\end{subfigure}	
\begin{subfigure}[t]{0.32\linewidth}
		\includegraphics[width=\linewidth]{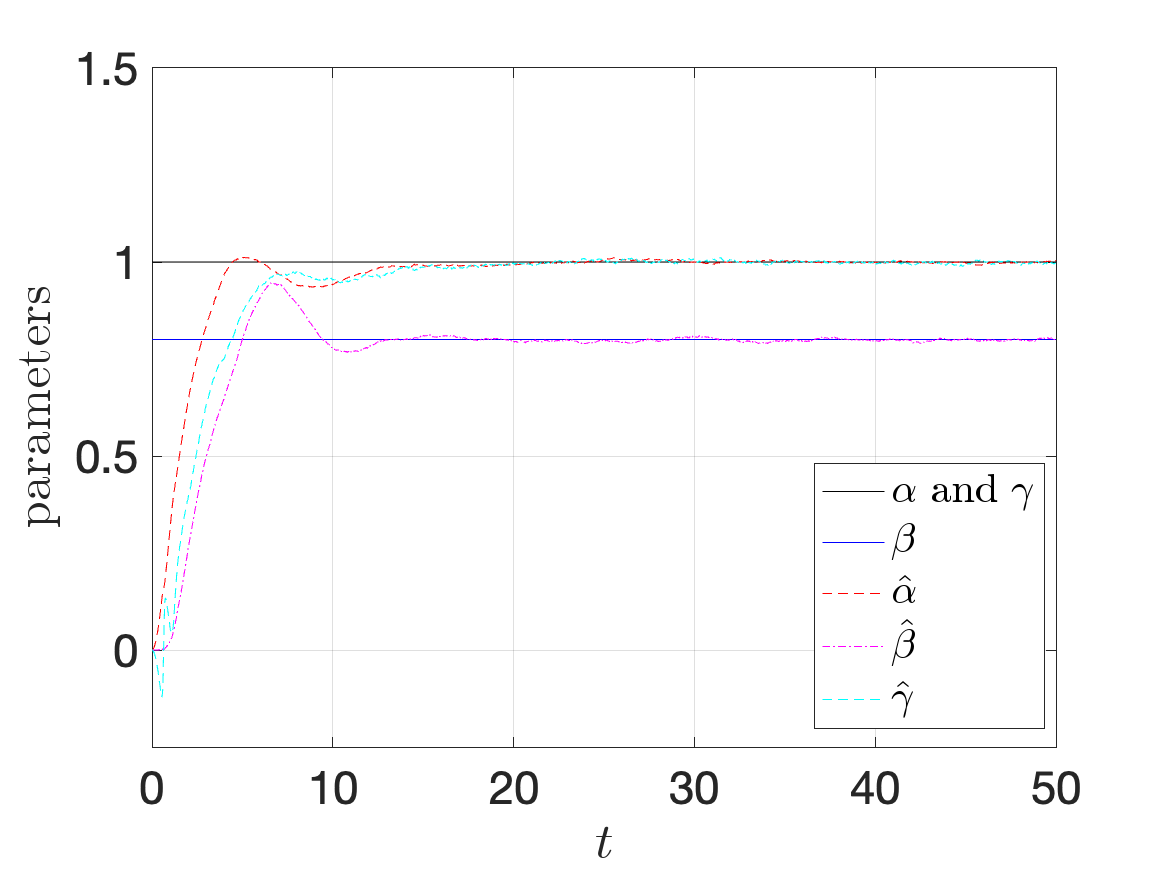}
		\caption{Parameter errors with $\beta=0.8$.}
		\label{f8i}
	\end{subfigure}	
}

\caption{\cblack{\textbf{(a, b, c)} Master signal with $\alpha=\gamma=1$ and $\beta=0$, $\beta=0.2$ and $\beta=0.8$, respectively. \textbf{(d, e, f)} Normalized synchronization MSE $\bar{\mathcal{E}}^2(t)$ versus time. \textbf{(g, h, i)} Parameter estimates versus time. The master and slave models are implmented with $M=K=64$ Fourier modes. The coupling and adaptation rate parameters in the slave model are $D=0.5$ and $\mu=200$, respectively. Observations are noisy, with average SNR$=12$~dB.}}
\label{f8regimes}
\end{figure}

\cblack{All the simulations presented so far correspond to the same set of parameter values $\bm{\theta} = [1.15, -0.05, 0.98]$, which yield a chaotic regime for Eq. \eqref{eqGenKS}. Figure \ref{f8regimes} shows some illustrative results for different parameter sets, leading to different regimes of the KS equation.}

\cblack{
As shown in \cite{Kudryashov21}, for fixed $\alpha=\gamma=1$, the dynamics of the KS equation shift from chaotic to periodic as the parameter $\beta$ is increased. Figures \ref{f8a}, \ref{f8b} and \ref{f8c} display the space-time plot of the field $u(t,x)$ for $T=100$ and $X=120$ when $\beta=0$, $\beta=0.2$ and $\beta=0.8$, respectively. We have run the synchronization-based method with $M=K=64$ for each of these three scenarios and we present the average normalized synchronization error $\bar{\mathcal{E}}^2(t)$ (Figures \ref{f8d}, \ref{f8e} and \ref{f8f}) and the evolution of the parameter estimates (Figures \ref{f8g}, \ref{f8h} and \ref{f8i}). We observe that convergence and accuracy are similar in all three cases (just slightly faster for $\beta=0.8$). 
}

\subsubsection{\cblack{Synchronization-based control}} \label{sssControl}

\cblack{
Synchronization techniques can often be used for the implementation of control schemes \cite{Zhang09} and the KS model has received specific attention from the control theory community \cite{Jamal15,Kang18}. For the proposed method, one can think of the slave model \eqref{eqParUpdate}-\eqref{eqCoeffUpdate} as a controlled system, where $u(t,x)$ is an input signal (i.e., not necessarily generated by a KS equation), $\bm{\theta}=[\alpha, \beta,\gamma]^\top$ is the control parameter and the aim is to make the slave signal $v_K(t,x)$ follow the reference input $u(t,x)$. 
}

\cblack{
We have carried out a simple computer simulation to test whether the slave model \eqref{eqParUpdate}-\eqref{eqCoeffUpdate} can be forced to follow an arbitrary input $u(x,t)$. This is just an illustrative experiment; a proper assessment of the proposed scheme in a control setup would require a detailed study.  
}

\begin{figure}
\centerline{	
	\begin{subfigure}[t]{0.43\linewidth}
		\includegraphics[width=\linewidth]{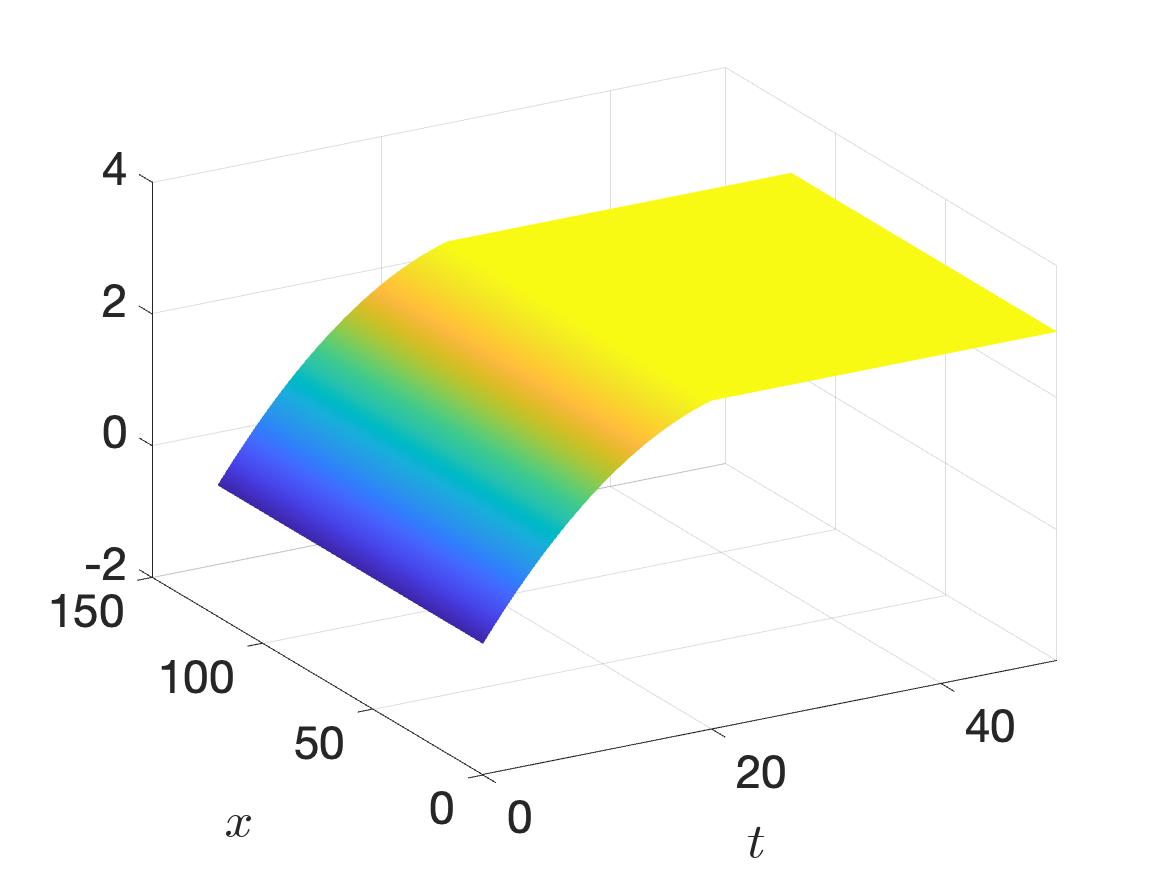}
		\caption{Reference signal (observations).}
		\label{f9a}
	\end{subfigure}
	\begin{subfigure}[t]{0.43\linewidth}
		\includegraphics[width=\linewidth]{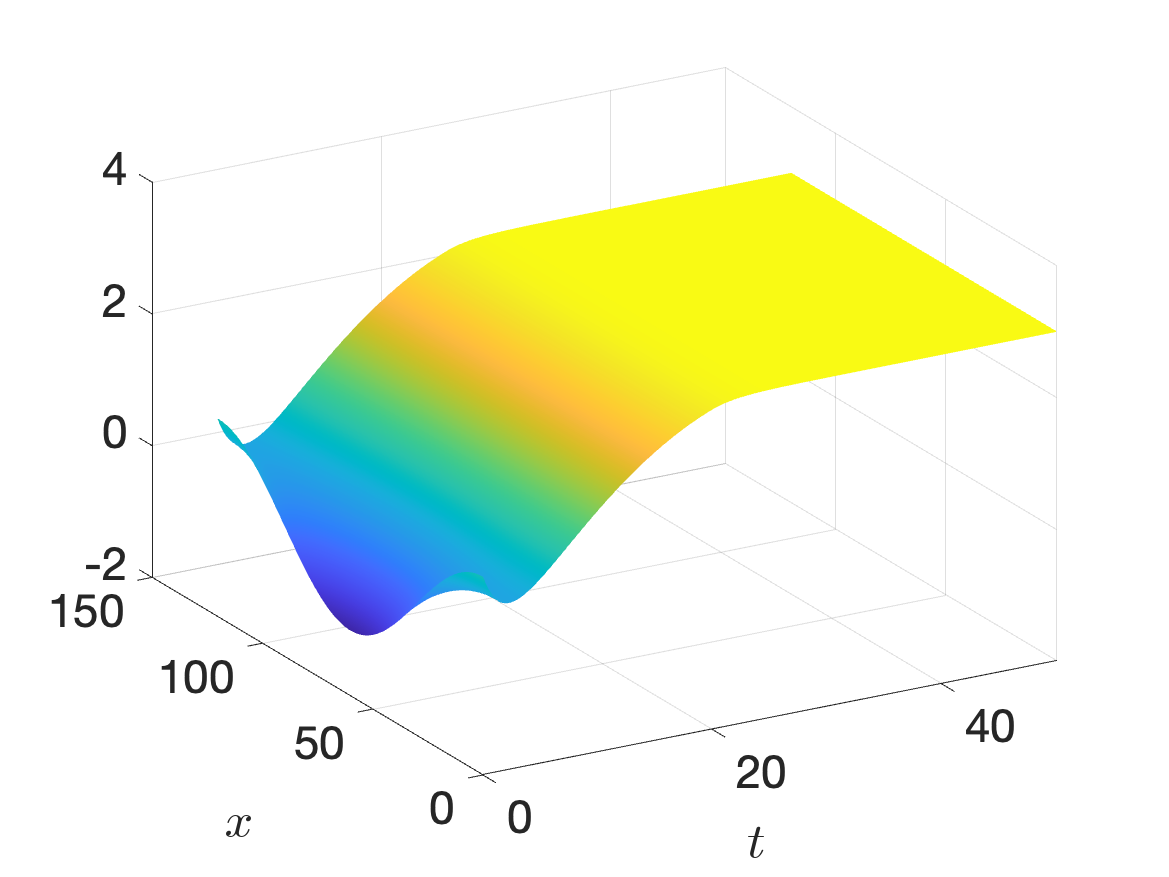}
		\caption{Slave model.}
		\label{f9b}
	\end{subfigure}	
}
\caption{\cblack{Control experiment. \textbf{(a)} Reference signal passed as observations to the slave model. \textbf{(b)} Slave model signal $u_K(x,t)$ with $K=64$. The coupling strength and adaptation rate are $D=0.5$ and $\mu=200$, respectively. The initial parameter values are $\hat{\bm{\theta}}(0)=[0,0,0]^\top$.}}
\label{f9control}
\end{figure}

\cblack{
The reference signal is depicted in Figure \ref{f9a}. It is initialised at $u(0,x) = 0$ and then increases smoothly to reach $u(20,x)=3$ for all $x \in [0, 120)$; then $u(t,x)=3$ for all $t \ge 20$. The parameter $\theta$ is initialised as $\hat{\bm{\theta}}=[0,0,0]^\top$ like in the previous experiments. Also, the coupling coefficient is $D=0.5$ and the adaptation rate is $\mu=200$. We set $K=64$ Fourier modes in the slave system.
}

\cblack{Figure \ref{f9b} shows the signal $v_K(t,x)$ of the slave model implemented with Eqs. \eqref{eqParUpdate}-\eqref{eqCoeffUpdate}. We observe how it reproduces the input signal accurately after a brief convergence period due to the mismatch in the initialization. Further study of the proposed scheme in more elaborate (and realistic) control problems is left for future work.}

%
\section{Conclusions} \label{sConclusions}

We have investigated the synchronization properties and the estimation of the constant parameters of a generalized Kuramoto-Sivashinsky (KS) equation in 1-dimensional space. We have assumed the ability to collect observations over time from a {\em master} system with possibly unknown parameters and then we have tackled the design of a {\em slave} model driven by the observations. We have proved that, when the parameters of the slave model are fixed and identical to the master parameters, the slave system attains local identical synchronization. This is ensured by Proposition \ref{propSynchro}, which is a relatively simple result yet, to our best knowledge, not available in the previous literature on the KS equation. When the master parameters are unknown, the parameters of the slave system are time-varying and driven by a (suitably designed) ODE that depends on the observations and the Fourier coefficients of the slave model. We have conducted a detailed numerical study and shown that synchronization and accurate parameter estimation can be achieved, \cblack{for different dynamical regimes (chaotic or periodic)}, even when the observations are noisy and the number of significant Fourier modes of the master system is underestimated.

The proposed scheme has turned out numerically robust to initialization errors (both in the signal and the parameters) in our simulations and it is designed to be implemented online and to estimate several parameters concurrently. \cblack{Most statistical methods that have been aplied to the KS model \cite{MartinaPerez21,Lu17,Huttunen18,Hurst22} are offline (i.e., observations are processed iteratively in batches, rather than sequentially as they are collected) and they are computationally much more demanding.
We have also compared the proposed scheme with the unscented Bucy-Kalman filter \cite{Sarkka07} (a popular {\em online} statistical estimation method for nonlinear systems) and found that the synchronization-based technique is more accurate and, even in this case, computationally lighter. However, the synchronization scheme requires continuous observation of the master system over a spatial grid (while most statistical schemes can work with discrete data sets) and does not provide a quantification of the expected error as Bayesian methods do. Compared to the scheme in \cite{Pachev22}, the proposed methodology can be robustly applied with noisy observations and is easier to implement. This is because it only demands the numerical integration of a relatively simple set of ODEs, while the method in \cite{Pachev22} requires the construction of an orthonormal basis and the solution of a system of linear equations at each integration step.} 

Future work includes a stability analysis of the slave model with time-varying parameters, \cblack{the combination of the proposed scheme with ensemble-based methods for data assimilation, the extension of the methodology to other nonlinear PDEs or the design of control schemes. A simple, preliminary example of a controlled KS model forced to follow an {\em ad hoc} input signal has been presented in Section \ref{sssControl}. Relevant nonlinear PDEs with a structure similar to Kuramoto-Sivashinsky's include, e.g., the Kawahara \cite{Topper78}, Benney-Lin \cite{Benney66,Lin74} or Nikolaevsky \cite{Simbawa10} equations. These PDEs are similar enough to the KS model that the proposed methodology can be applied along the same lines described in this paper. Additional research will be needed to extend the stability analysis (possibly to a class of PDEs including these specific equations) and assess the numerical performance of the resulting schemes.}

\begin{acknowledgments}
This work has been partially supported by the Office of Naval Research (award N00014-22-1-2647) and Spain's {\em Agencia Estatal de Investigaci\'on} (ref. PID2021-125159NB-I00 TYCHE) funded by MCIN/AEI/10.13039/501100011033 and by ``ERDF A way of making Europe".
\end{acknowledgments}

%
\section*{Data Availability}


The data that support the findings of this study are available within the article and its supplementary material.

%

%
\bibliographystyle{unsrt}
\bibliography{bibliografia_JM}

\end{document}